\documentclass[longauth]{aa}

\usepackage{arydshln}
\usepackage{comment}
\usepackage{graphicx}
\usepackage{multirow}
\usepackage{pifont}
\usepackage[normalem]{ulem}
\usepackage{xcolor}
\usepackage[varg]{txfonts}
\usepackage[colorlinks=true,citecolor=blue, urlcolor=blue, linkcolor=blue]{hyperref}

\makeatletter
\renewcommand*\aa@pageof{, page \thepage{} of \pageref*{LastPage}}
\makeatother

\newcommand{\micron}{\text{\textmu m}}
\newcommand{\red}[1]{\textcolor{black}{#1}}

\begin{document}

   \title{The ExoGRAVITY survey: a \textit{K}-band spectral library of giant exoplanet and brown dwarf companions}

\author{J.~Kammerer\inst{\ref{esog}}
 \and T.~O.~Winterhalder\inst{\ref{esog}}
 \and S.~Lacour\inst{\ref{lesia},\ref{esog}}
 \and T.~Stolker\inst{\ref{leiden}}
 \and G.-D.~Marleau\inst{\ref{bernWP},\ref{mpia},\ref{duisburg}}
 \and W.~O.~Balmer\inst{\ref{jhupa},\ref{stsci}}
 \and A.~F.~Moore\inst{\ref{ucc}}
 \and L.~Piscarreta\inst{\ref{esog}}
 \and C.~Toci\inst{\ref{esog}}
 \and A.~M\'erand\inst{\ref{esog}}
 %\and \textcolor{red}{M.~Nowak\inst{\ref{lesia}}}
 \and M.~Nowak\inst{\ref{lesia}}
 \and E.~L.~Rickman\inst{\ref{esa}}
 \and L.~Pueyo\inst{\ref{stsci}}
 \and N.~Pourr\'e\inst{\ref{ipag}}
 \and E.~Nasedkin\inst{\ref{tnt}}
 \and J.~J.~Wang\inst{\ref{northwestern}}
 \and G.~Bourdarot\inst{\ref{mpe}}
 \and F.~Eisenhauer\inst{\ref{mpe}}
 \and Th.~Henning\inst{\ref{mpia}}
 \and R.~Garcia~Lopez\inst{\ref{dublin},\ref{mpia}}
 \and E.F.~van~Dishoeck\inst{\ref{leiden},\ref{mpe}}
 \and T.~Forveille\inst{\ref{ipag}}
 \and J.~D.~Monnier\inst{\ref{umich}}
 \and R.~Abuter\inst{\ref{esog}}
 \and A.~Amorim\inst{\ref{lisboa},\ref{centra}}
 %\and \textcolor{red}{M.~Benisty\inst{\ref{mpia},\ref{ipag}}}
 \and M.~Benisty\inst{\ref{mpia},\ref{ipag}}
 \and J.-P.~Berger\inst{\ref{ipag}}
 \and H.~Beust\inst{\ref{ipag}}
 \and S.~Blunt\inst{\ref{northwestern}}
 \and A.~Boccaletti\inst{\ref{lesia}}
 \and M.~Bonnefoy\inst{\ref{ipag}}
 \and H.~Bonnet\inst{\ref{esog}}
 \and M.~S.~Bordoni\inst{\ref{mpe}}
 \and W.~Brandner\inst{\ref{mpia}}
 \and F.~Cantalloube\inst{\ref{ipag}}
 \and P.~Caselli\inst{\ref{mpe}}
 \and W.~Ceva\inst{\ref{geneva}}
 \and B.~Charnay\inst{\ref{lesia}}
 \and G.~Chauvin\inst{\ref{cotedazur}}
 \and A.~Chavez\inst{\ref{northwestern}}
 \and A.~Chomez\inst{\ref{lesia},\ref{ipag}}
 \and E.~Choquet\inst{\ref{lam}}
 \and V.~Christiaens\inst{\ref{liege}}
 \and Y.~Cl\'enet\inst{\ref{lesia}}
 \and V.~Coud\'e~du~Foresto\inst{\ref{lesia}}
 \and A.~Cridland\inst{\ref{leiden}}
 \and R.~Davies\inst{\ref{mpe}}
 \and R.~Dembet\inst{\ref{lesia}}
 \and J.~Dexter\inst{\ref{boulder}}
 \and A.~Drescher\inst{\ref{mpe}}
 \and G.~Duvert\inst{\ref{ipag}}
 \and A.~Eckart\inst{\ref{cologne},\ref{bonn}}
 \and C.~Fontanive\inst{\ref{montreal}}
 \and N.~M.~F\"orster Schreiber\inst{\ref{mpe}}
 \and P.~Garcia\inst{\ref{centra},\ref{porto}}
 \and E.~Gendron\inst{\ref{lesia}}
 \and R.~Genzel\inst{\ref{mpe},\ref{ucb}}
 \and S.~Gillessen\inst{\ref{mpe}}
 \and J.~H.~Girard\inst{\ref{stsci}}
 \and S.~Grant\inst{\ref{mpe}}
 \and J.~Hagelberg\inst{\ref{geneva}}
 \and X.~Haubois\inst{\ref{esoc}}
 \and G.~Hei\ss el\inst{\ref{actesa},\ref{lesia}}
 \and S.~Hinkley\inst{\ref{exeter}}
 \and S.~Hippler\inst{\ref{mpia}}
 \and M.~Houll\'e\inst{\ref{cotedazur}}
 \and Z.~Hubert\inst{\ref{ipag}}
 \and L.~Jocou\inst{\ref{ipag}}
 \and M.~Keppler\inst{\ref{mpia}}
 \and P.~Kervella\inst{\ref{lesia}}
 \and L.~Kreidberg\inst{\ref{mpia}}
 \and N.~T.~Kurtovic\inst{\ref{mpe}}
 \and A.-M.~Lagrange\inst{\ref{ipag},\ref{lesia}}
 \and V.~Lapeyr\`ere\inst{\ref{lesia}}
 \and J.-B.~Le~Bouquin\inst{\ref{ipag}}
 \and D.~Lutz\inst{\ref{mpe}}
 \and A.-L.~Maire\inst{\ref{ipag}}
 \and F.~Mang\inst{\ref{mpe}}
 \and E.~C.~Matthews\inst{\ref{mpia}}
 \and P.~Molli\`ere\inst{\ref{mpia}}
 \and C.~Mordasini\inst{\ref{bernWP},\ref{bernCSH}}
 \and D.~Mouillet\inst{\ref{ipag}}
 \and T.~Ott\inst{\ref{mpe}}
 \and G.~P.~P.~L.~Otten\inst{\ref{sinica}}
 \and C.~Paladini\inst{\ref{esoc}}
 \and T.~Paumard\inst{\ref{lesia}}
 \and K.~Perraut\inst{\ref{ipag}}
 \and G.~Perrin\inst{\ref{lesia}}
 \and O.~Pfuhl\inst{\ref{esog}}
 \and D.~C.~Ribeiro\inst{\ref{mpe}}
 \and Z.~Rustamkulov\inst{\ref{jhueps}}
 \and D.~S\'egransan\inst{\ref{geneva}}
 \and J.~Shangguan\inst{\ref{beijing}}
 \and T.~Shimizu\inst{\ref{mpe}}
 \and M.~Samland\inst{\ref{mpia}}
 \and D.~Sing\inst{\ref{jhupa},\ref{jhueps}}
 \and J.~Stadler\inst{\ref{mpa},\ref{origins}}
 \and O.~Straub\inst{\ref{origins}}
 \and C.~Straubmeier\inst{\ref{cologne}}
 \and E.~Sturm\inst{\ref{mpe}}
 \and L.~J.~Tacconi\inst{\ref{mpe}}
 \and S.~Udry\inst{\ref{geneva}}
 \and A.~Vigan\inst{\ref{lam}}
 \and F.~Vincent\inst{\ref{lesia}}
 \and S.~D.~von~Fellenberg\inst{\ref{bonn}}
 \and F.~Widmann\inst{\ref{mpe}}
 \and J.~Woillez\inst{\ref{esog}}
 \and S.~Yazici\inst{\ref{mpe}}
 \and the GRAVITY Collaboration}

\institute{ 
   European Southern Observatory, Karl-Schwarzschild-Stra\ss e 2, 85748 Garching, Germany\\
   \email{jkammere@eso.org}
\label{esog}      \and
   LESIA, Observatoire de Paris, PSL, CNRS, Sorbonne Universit\'e, Universit\'e de Paris, 5 place Janssen, 92195 Meudon, France
\label{lesia}      \and
   Leiden Observatory, Leiden University, Einsteinweg 55, 2333 CC Leiden, The Netherlands
\label{leiden}      \and
   Division of Space Research \&\ Planetary Sciences, Physics Institute, University of Bern, Sidlerstr.~5, 3012 Bern, Switzerland
\label{bernWP}        \and
    Max-Planck-Institut f\"ur Astronomie, K\"onigstuhl 17, 69117 Heidelberg, Germany
\label{mpia}  \and
    Fakult\"at für Physik, Universit\"at Duisburg--Essen, Lotharstra\ss{}e 1, 47057 Duisburg, Germany
\label{duisburg}      \and
   Department of Physics \& Astronomy, Johns Hopkins University, 3400 N. Charles Street, Baltimore, MD 21218, USA
\label{jhupa}      \and
   Space Telescope Science Institute, 3700 San Martin Drive, Baltimore, MD 21218, USA
\label{stsci}      \and
   Department of Physics, University College Cork, Cork, Ireland
\label{ucc}      \and
   European Space Agency (ESA), ESA Office, Space Telescope Science Institute, 3700 San Martin Drive, Baltimore, MD 21218, USA
\label{esa}      \and
   Univ. Grenoble Alpes, CNRS, IPAG, 38000 Grenoble, France
\label{ipag}      \and
   School of Physics, Trinity College Dublin, The University of Dublin, Dublin 2, Ireland
\label{tnt}    \and
   Center for Interdisciplinary Exploration and Research in Astrophysics (CIERA) and Department of Physics and Astronomy, Northwestern University, Evanston, IL 60208, USA
\label{northwestern}      \and
   Max Planck Institute for extraterrestrial Physics, Giessenbachstra\ss e~1, 85748 Garching, Germany
\label{mpe}      \and
   School of Physics, University College Dublin, Belfield, Dublin 4, Ireland
\label{dublin}      \and
   Astronomy Department, University of Michigan, Ann Arbor, MI 48109 USA
\label{umich}      \and
   Universidade de Lisboa - Faculdade de Ci\^encias, Campo Grande, 1749-016 Lisboa, Portugal
\label{lisboa}      \and
   CENTRA - Centro de Astrof\' isica e Gravita\c c\~ao, IST, Universidade de Lisboa, 1049-001 Lisboa, Portugal
\label{centra}      \and
   Départment d’astronomie de l’Université de Genève, 51 ch. des Maillettes Sauverny, 1290 Versoix, Switzerland
\label{geneva}      \and
   Université Côte d’Azur, Observatoire de la Côte d’Azur, CNRS, Laboratoire Lagrange, Bd de l'Observatoire, CS 34229, 06304 Nice cedex 4, France
\label{cotedazur}      \and
   Aix Marseille Univ, CNRS, CNES, LAM, Marseille, France
\label{lam}      \and
  STAR Institute, Universit\'e de Li\`ege, All\'ee du Six Ao\^ut 19c, 4000 Li\`ege, Belgium
\label{liege}      \and
   Department of Astrophysical \& Planetary Sciences, JILA, Duane Physics Bldg., 2000 Colorado Ave, University of Colorado, Boulder, CO 80309, USA
\label{boulder}      \and
   1. Institute of Physics, University of Cologne, Z\"ulpicher Stra\ss e 77, 50937 Cologne, Germany
\label{cologne}      \and
   Max Planck Institute for Radio Astronomy, Auf dem H\"ugel 69, 53121 Bonn, Germany
\label{bonn}      \and
   Trottier Institute for Research on Exoplanets, Université de Montreal, Montreal, Quebec H3C 3J7, Canada
\label{montreal}
   Universidade do Porto, Faculdade de Engenharia, Rua Dr.~Roberto Frias, 4200-465 Porto, Portugal
\label{porto}      \and
   Departments of Physics and Astronomy, Le Conte Hall, University of California, Berkeley, CA 94720, USA
\label{ucb}      \and
   European Southern Observatory, Casilla 19001, Santiago 19, Chile
\label{esoc}      \and
   Advanced Concepts Team, European Space Agency, TEC-SF, ESTEC, Keplerlaan 1, NL-2201, AZ Noordwijk, The Netherlands
\label{actesa}      \and
   University of Exeter, Physics Building, Stocker Road, Exeter EX4 4QL, United Kingdom
\label{exeter}      \and
   Center for Space and Habitability, University of Bern, Gesellschaftsstr.~6, 3012 Bern, Switzerland
\label{bernCSH}    \and
   Academia Sinica, Institute of Astronomy and Astrophysics, 11F Astronomy-Mathematics Building, NTU/AS campus, No. 1, Section 4, Roosevelt Rd., Taipei 10617, Taiwan
\label{sinica}      \and
   Department of Earth \& Planetary Sciences, Johns Hopkins University, Baltimore, MD, USA
\label{jhueps}      \and
   The Kavli Institute for Astronomy and Astrophysics, Peking University, Beijing 100871, China
\label{beijing}      \and
   Max Planck Institute for Astrophysics, Karl-Schwarzschild-Str. 1, 85741 Garching, Germany
\label{mpa}      \and
   Excellence Cluster ORIGINS, Boltzmannstraße 2, D-85748 Garching bei München, Germany
\label{origins}
}

   \date{Received August 15, 2025; accepted October 8, 2025}

% \abstract{}{}{}{}{} 
% 5 {} token are mandatory
 
  \abstract
  % context heading (optional)
  % {} leave it empty if necessary  
   {Direct observations of exoplanet and brown dwarf companions with near-infrared interferometry, first enabled by the dual-field mode of VLTI/GRAVITY, provide unique measurements of the objects' orbital motion and atmospheric composition.}
  % aims heading (mandatory)
   {Here, we compile a homogeneous library of all exoplanet and brown dwarf \textit{K}-band spectra observed by GRAVITY thus far. This ExoGRAVITY Spectral Library is made publicly available online.}
  % methods heading (mandatory)
   {\red{We re-reduce all available GRAVITY dual-field high-contrast data in a uniform and highly automated way and, where companions are detected, extract their $\sim2.0$--$2.4~\micron$ \textit{K}-band contrast spectra.} We then derive stellar model atmospheres for all employed flux references (either the host star or the swap calibrator) that we use to convert the companion contrast into companion flux spectra. Solely from the resulting GRAVITY \textit{K}-band flux spectra, we extract spectral types, spectral indices, and bulk physical properties for all companions. Finally, and with the help of age constraints from the literature, we also derive isochronal masses for most companions using evolutionary models.}
  % results heading (mandatory)
   {The resulting library contains $R \sim 500$ GRAVITY \textit{K}-band spectra of 39 substellar companions from late M to late T spectral type, including the entire L-T transition. Throughout this transition, a shift from CO-dominated late M- and L-type dwarfs to CH${}_4$-dominated T-type dwarfs can be observed in the \textit{K}-band. The GRAVITY spectra alone constrain the objects' bolometric luminosity to typically within $\pm0.15$~dex. The derived isochronal masses agree with dynamical masses from the literature where available, except for HD~4113~c for which we confirm its previously reported potential underluminosity.}
  % conclusions heading (optional), leave it empty if necessary 
   {Medium resolution spectroscopy of substellar companions with GRAVITY provides insight into the carbon chemistry and the cloudiness of these objects' atmospheres. It also constrains these objects' bolometric luminosity which can yield measurements of their formation entropy if combined with dynamical masses, for instance from Gaia and GRAVITY astrometry.}


   \maketitle
%
%-------------------------------------------------------------------

\section{Introduction}
\label{sec:introduction}

Direct observations of substellar companions have proven powerful at studying the population of young and self-luminous exoplanets and brown dwarfs \citep[e.g.,][]{bowler2016,currie2023b}. Direct detection techniques provide access to companions at wide separations from the star which are difficult to observe through transits or radial velocities. This makes direct observations ideally suited to study the population of gas giant planets and brown dwarfs at tens to hundreds of au from the host star and to unravel their formation and early evolution history. Spectroscopic observations at low to high spectral resolution provide insight into the chemical composition and atmospheric dynamics of these companions. This enables constraining elemental and isotopic abundance ratios \citep[e.g.,][]{hoch2023,zhang2021} which are thought to be linked to the companions' formation environments and mechanisms \citep[e.g.,][]{oeberg2011,eistrup2018,molliere2022}. Ultimately, direct detection techniques are poised to unravel the origin and dynamical evolution of gas giant planets and to help constrain different planet formation scenarios such as core accretion \citep{pollack1996}, disk instability \citep{boss1997}, and cloud fragmentation \citep{bate2012}.

Classical high-contrast imaging instruments on 8--10~m-class telescopes have been limited to detecting companions at $>10$~au separations from the host star due to the inner working angle of their coronagraphs \citep[e.g.,][]{nielsen2019,vigan2021}. Interferometric techniques are one avenue to overcome this limitation. While single-telescope aperture masking and kernel phase interferometry from the ground are struggling to achieve the necessary contrasts to detect planetary-mass companions \citep[e.g.,][]{ireland2013,kammerer2019,wallace2020,stolker2024}, the Near Infrared Imager and Slitless Spectrograph (NIRISS) instrument \citep{doyon2023} onboard \emph{JWST} \citep{gardner2023} is capable of imaging exoplanetary companions down to $\sim70$~mas separations \red{using Aperture Masking Interferometry with the non-redundant mask} (\citealt{sivaramakrishnan2023}, \citealt{desdoigts2025}), despite only in broadband photometry. With the detection of the exoplanet HR~8799~e with VLTI/GRAVITY \citep{gravity2019}, long-baseline interferometry entered the field of direct exoplanet detection techniques and opened up a new parameter space for direct observations of substellar companions at $\sim1$--10~au separations from the host star. The unique combination of angular resolution and starlight suppression capabilities with long-baseline interferometry enables measuring companion astrometry at $\sim50~\text{\textmu as}$ precision and companion spectra at a high signal-to-noise ratio \citep[SNR,][]{lacour2020}. GRAVITY in particular provides spectroscopy at $\sim2.0$--$2.4~\micron$ wavelength in the \textit{K}-band which is a powerful spectral regime for studying the atmospheres of substellar companions near the L-T transition. Molecular features from carbon-monoxide (CO) and methane (CH${}_4$) help constraining the carbon chemistry and cloudiness of these objects \citep[e.g.,][]{patience2012,zahnle2014} and determining elemental abundance ratios, in particular C/O ratios \citep{gravity2020,molliere2020}. These have been discussed in the literature for constraining formation scenarios \citep[e.g.,][]{oeberg2011,eistrup2018,molliere2022,hoch2023}.

Over the past years, GRAVITY has been used to observe a large fraction of the currently known sample of directly-imaged exoplanet and brown dwarf companions. Pioneering work by \citet{gravity2019,gravity2020} demonstrated the depth at which GRAVITY can study the atmospheres of nearby gas giant exoplanets. This work laid the foundation for the ExoGRAVITY Collaboration, which was established to broaden the team beyond the original GRAVITY Collaboration, disseminate technical knowledge, and foster collaboration with the wider exoplanet community. Within this ExoGRAVITY effort,
\citet{wang2021} showed that GRAVITY is not far from being able to spatially resolve the circumplanetary environment of the accreting PDS~70 planets and that these two planets are close to being in a 2:1 mean motion resonance. \citet{lagrange2020} and \citet{kammerer2021} paved the way for the first direct detection of $\beta$~Pic~c at $\sim2.7$~au separation \citep{nowak2020} and HD~206893~c at $\sim3.5$~au separation \citep{hinkley2023}, only made possible thanks to GRAVITY's unparalleled spatial resolution. Various further studies demonstrated the characterization potential of the GRAVITY \textit{K}-band spectra if combined with spectrophotometry at other wavelengths from the literature \citep{balmer2023,balmer2024,balmer2025,winterhalder2025,stolker2025}.

With the advent of Gaia Data Release 4 (DR4), GRAVITY will be capable of following up dozens of astrometrically detected and young exoplanets at $\sim1$--10~au separations from the host star where currently no other instrument can characterize their atmospheres \citep{perryman2014}. \citet{winterhalder2024} have shown that a single GRAVITY epoch combined with the orbital parameters from Gaia can yield precise dynamical masses for substellar companions so that studies of their formation entropies might be possible if their bolometric luminosities can be sufficiently constrained with GRAVITY \citep[e.g.,][]{mordasini2017,marleau2019}. Currently, GRAVITY is undergoing the GRAVITY+ upgrade which further increases the spatial resolution and SNR at which GRAVITY can observe planetary-mass companions \citep{gravity2025} and enhances the prospects for Gaia follow-up observations of exoplanets. \red{With the new adaptive optics system, companions down to $\sim21$~apparent magnitude in the \textit{K}-band \citep[such as HD~4113~c,][]{cheetham2018} are already in reach.}

In this paper, we derive and analyze the GRAVITY \textit{K}-band spectra of the currently 39 substellar companions whose data is either publicly available or part of the ExoGRAVITY Collaboration \citep{lacour2020}. There will be a separate publication by Roberts et al. (in prep.) dealing with the measured GRAVITY astrometry of all these objects. During the remainder of this paper, we denote all companions with lowercase letters (i.e., b/c/d/e) regardless of their nature (e.g., giant planet or brown dwarf). This choice was made because in this paper, we do not explicitly differentiate between giant planets and brown dwarfs in any of our analyses and lowercase letters are preferential for distinguishing substellar companions from stellar multiples which we denote with uppercase letters. \red{This paper presents a basic atmospheric characterization of the companions, including an estimation of evolutionary masses and a comparison to dynamical masses from the literature where available. However, this analysis is done using only the GRAVITY \textit{K}-band spectra. At no point in this paper, we are drawing any conclusions regarding the nature or formation history of the companions as this would require a more detailed analysis considering data over a broader spectral range which is left for future work.}

The paper is organized as follows. Section~\ref{sec:observations_and_data_reduction} summarizes the GRAVITY observations and explains the data reduction procedure as well as some details about the data format of the online spectral library that we are providing. Section~\ref{sec:results} presents the GRAVITY \textit{K}-band spectra of the 39 substellar companions in the library including some basic spectral type and atmospheric characterization. In Section~\ref{sec:discussion} we derive isochronal masses for most companions using age constraints from the literature where available. Finally, Section~\ref{sec:summary_and_conclusions} summarizes our results and concludes with our main findings from this paper.

\section{Observations and data reduction}
\label{sec:observations_and_data_reduction}

\subsection{GRAVITY observations}
\label{sec:gravity_observations}

All exoplanet and brown dwarf companion spectra presented in this work were obtained with the GRAVITY instrument \citep{gravity2017} at the Very Large Telescope Interferometer (VLTI) at the European Southern Observatory's (ESO's) Paranal Observatory in Chile. The observational setup, integration time, ambient conditions, and corresponding program ID and PI for each dataset and object can be found on Zenodo\footnote{Available at \url{https://doi.org/10.5281/zenodo.17295254}}. Epochs which are crossed out were discarded because the derived spectra suffered from strong systematics (see also Section~\ref{sec:systematics_in_the_gravity_spectra}).

All observations were done in dual-field mode following the strategy outlined in \citet{gravity2019}. Close-in companions were typically observed on-axis, meaning with the 50/50 beam splitter to direct half of the light of the host star into the fringe tracker and half of the light of the companion into the science spectrometer, set to a spectral resolution of $R \sim 500$. The dual-field on-axis mode is suitable for companions with an angular separation of less than \red{$\sim600$~mas with the 8.2~m Unit Telescopes (UTs) and $\sim2.7\arcsec$ with the 1.8~m Auxiliary Telescopes (ATs)} from the host star. Due to the use of the 50/50 beamsplitter, the efficiency of this mode is only 50\%. Companions which are further away from the host star or which are closer than 600~mas/2.7\arcsec but very faint were typically observed off-axis, meaning with the roof prism to direct 100\% of the light of the companion into the science spectrometer. The dual-field off-axis mode is suitable for companions with angular separations between 270~mas and $2\arcsec$ with the UTs (1.17--4\arcsec with the ATs) from the host star.

The obvious advantage of dual-field off-axis observations is the better throughput. However, the advantage of dual-field on-axis observations is that the science spectrometer observations of the companion can be interspersed with short observations of the host star to obtain a quasi-simultaneous contrast calibration of the companion spectrum with respect to the host star spectrum. This is not possible for dual-field off-axis observations because the roof prism splits the observed field into two separate parts and the science spectrometer can only observe the part where the companion, but not the part where the host star is located. In some cases, a dedicated on-axis observation of the host star was done before or after the off-axis observation of the companion to enable contrast calibration of the companion spectrum with respect to the host star spectrum. \red{Compared to on-axis observations}, this calibration is not quasi-simultaneous and typically taken $\sim$half an hour apart from the companion observation. Going forward, we highly recommend to always take such a dedicated on-axis host star observations for all off-axis companion observations since it substantially improves the flux calibration of the resulting companion spectrum. If such a dedicated on-axis observation of the host star is not available, the companion spectrum can only be calibrated relative to the swap (binary star) calibrator that was observed before or after the off-axis observation of the companion. The primary purpose of the swap calibrator observation is to measure the optical path difference (OPD) between the two fields created by the roof prism. This is necessary to derive precise companion astrometry with respect to the host star. However, the swap calibrator observation can also be used to flux-calibrate the companion spectrum as described in Appendix~\ref{sec:flux_calibration_with_off-axis_binary_star_reference}.

\subsection{GRAVITY data reduction}
\label{sec:gravity_data_reduction}

All GRAVITY data were reduced with the publicly available consortium Python tools\footnote{Available at \url{https://version-lesia.obspm.fr/repos/DRS_gravity/gravi_tools3/}} which employ the official ESO Reflex scripts. We first used the \texttt{run\_gravi\_reduce} Python script to reduce the raw images to the so-called \texttt{astroreduced} files using version 1.6.4b1 of the ESO GRAVITY pipeline. Then, we used the \texttt{run\_gravi\_astrored\_astrometry} Python script to further process the \texttt{astroreduced} files and extract the \red{best fit} companion relative astrometry and contrast spectra. This reduction step which applies to GRAVITY observations of substellar companions in dual-field mode only follows the general procedure outlined in Appendix~A of \citet{gravity2020}. \red{This work only presents the cases where a significant companion could be detected in the GRAVITY data. There are several GRAVITY observations of candidate companions which did not yield significant detections for instance. However, the scientific interpretation of these non-detections is left for another publication.} We also corrected for fiber coupling losses arising from pointing offsets according to Appendix~A of \citet{wang2021}. For the vast majority of the observed targets, we removed the residual stellar light in the companion pointing using a model consisting of a 6th order complex polynomial. This model was fitted to the visibilities measured in the on-star pointing, and then subtracted from the on-companion pointing after phase shifting it to the companion fiber position. For three targets which are all located at close angular separation from the host star (BD+70~260~b, HD~17155~b, and HD~135344~Ab) we instead used a polynomial of 3rd order only to avoid overfitting and oversubtracting the companion spectrum. The exact angular separation at which using a lower order polynomial to remove the residual stellar light becomes preferable over a higher order one depends on the observing conditions and the brightness of the companion.

\subsection{Flux calibration using stellar model atmospheres}
\label{sec:flux_calibration_using_stellar_model_atmospheres}

To convert the companion contrast spectra measured by GRAVITY into companion flux spectra, a spectrum of the respective flux reference source is needed. For dual-field on-axis observations, \red{this is the host star}. For dual-field off-axis observations where no on-axis exposure on the host star is available, the companion contrast spectra were calibrated with respect to the swap (binary star) calibrator. The methodology for datasets where an on-axis exposure on the host star is available is described in detail in Appendix~\ref{sec:flux_calibration_with_on-axis_host_star_reference}. The differences and additional steps involved when using the swap calibrator for the flux calibration are explained in Appendix~\ref{sec:flux_calibration_with_off-axis_binary_star_reference}. In short, to obtain the final companion flux spectra $F_\text{comp}$, the companion contrast spectra $c_\text{comp}$ measured by GRAVITY were multiplied with the host star model spectra $F_\text{star}$ evaluated at the same wavelength nodes $\lambda$ as the GRAVITY spectra, that is
\begin{align}
    F_\text{comp}(\lambda) = c_\text{comp}(\lambda) \cdot F_\text{star}(\lambda).
\end{align}
When using the swap calibrator for the flux calibration, the term $F_\text{star}$ takes a more complex form which can be found in Appendix~\ref{sec:flux_calibration_with_off-axis_binary_star_reference}.

\begin{figure}
    \centering
    \includegraphics[trim={0.1cm 0.1cm 0 0},clip,width=\linewidth]{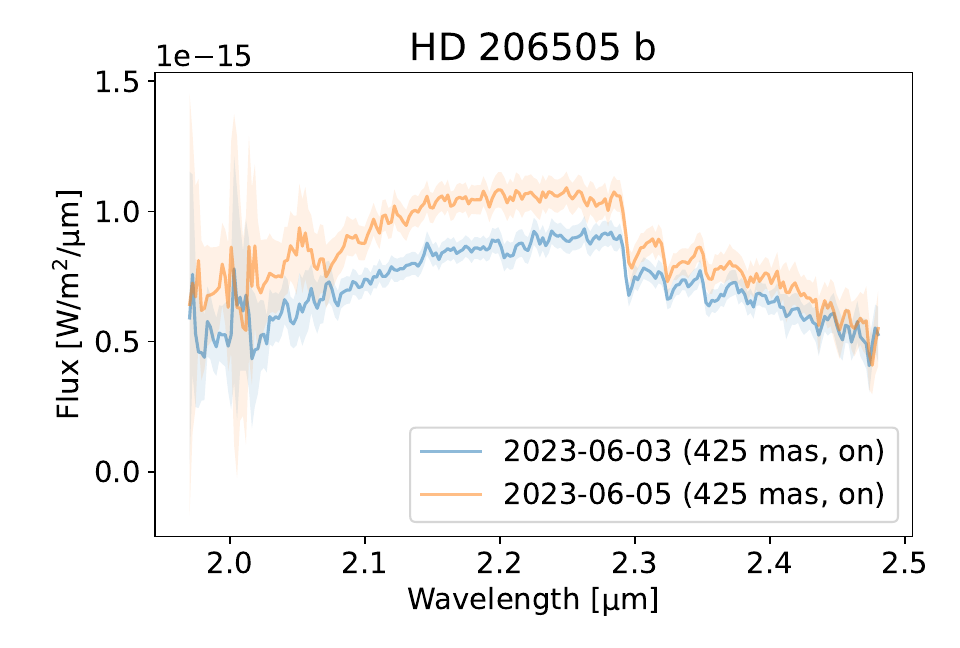}
    \includegraphics[trim={0.1cm 0.1cm 0 0},clip,width=\linewidth]{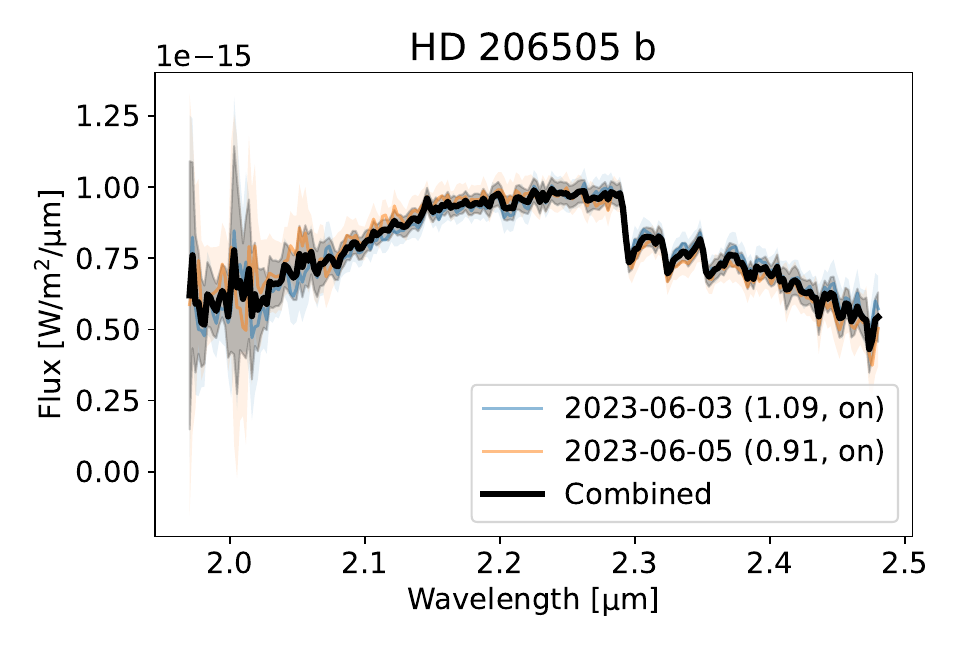}
    \caption{\red{Combination of the GRAVITY epochs of HD~206505~b after scaling them to bring them into mutual agreement.} \textit{Top panel}: GRAVITY \textit{K}-band flux spectra for each individual epoch. The $3\sigma$ flux uncertainties are shown as shaded regions. The legend reveals that for this object, both observations were done in dual-field on-axis mode, with the companion located at an angular separation of $425$~mas from the host star. \textit{Bottom panel}: Scaled GRAVITY flux spectra for each individual epoch and the resulting combined spectrum (black line). The legend reveals the scale factors applied to each epoch (values in brackets).}
    \label{fig:hd206505_b_all}
\end{figure}

\subsection{Combining spectra from multiple epochs}
\label{sec:combining_spectra_from_multiple_epochs}

To maximize the SNR, we combined all available spectra for a given object. However, before companion spectra from multiple epochs can be combined, it needs to be acknowledged that the calibration of the GRAVITY contrast spectra is affected by significant systematic errors. This results in companion spectra of the same object observed at different epochs being discrepant by more than what is expected based on their statistical error bars. An example for this is shown in Figure~\ref{fig:hd206505_b_all} (top panel), where the HD~206505~b spectrum observed on 3 June 2023 falls significantly below the one observed on 5 June 2023. This is a rather drastic example where the individual epochs disagree by $\sim18\%$ in flux. We emphasize that all of this discrepancy is inherent to the GRAVITY contrast spectra and has nothing to do with our stellar model atmosphere of the host star used to convert the companion contrast into a companion flux spectrum (since the same stellar model atmosphere was used to flux-calibrate both epochs).

\red{Figure~\ref{fig:hd206505_b_all} illustrates that scaling the individual GRAVITY epochs before combining them is necessary, and that the scaled epochs agree with each other fairly well. Unfortunately, this scaling process also removes any astrophysical variability from the data which is a major shortcoming of our library. However, there is currently no way of disentangling systematic (instrumental) from intrinsic (astrophysical) variability for the GRAVITY spectra. In case such methods are developed in the future, we also provide the unscaled GRAVITY spectra of all individual epochs in the online library (see Section~\ref{sec:description_of_data_format}). We finally note that the intrinsic variability of brown dwarfs \citep{vos2019,vos2020} and giant exoplanets \citep{biller2021,wang2022} is typically lower than $\sim5\%$ with the exception of a few outliers, so that we suspect that the flux variations of typically $\pm10\%$ seen for individual objects in many of the GRAVITY spectra are mainly of instrumental cause. \red{Potential origins} of such large systematic flux variations in the GRAVITY contrast spectra are discussed in Section~\ref{sec:systematics_in_the_gravity_spectra}.}

\begin{figure}
    \centering
    \includegraphics[trim={0.1cm 0.1cm 0 0},clip,width=\linewidth]{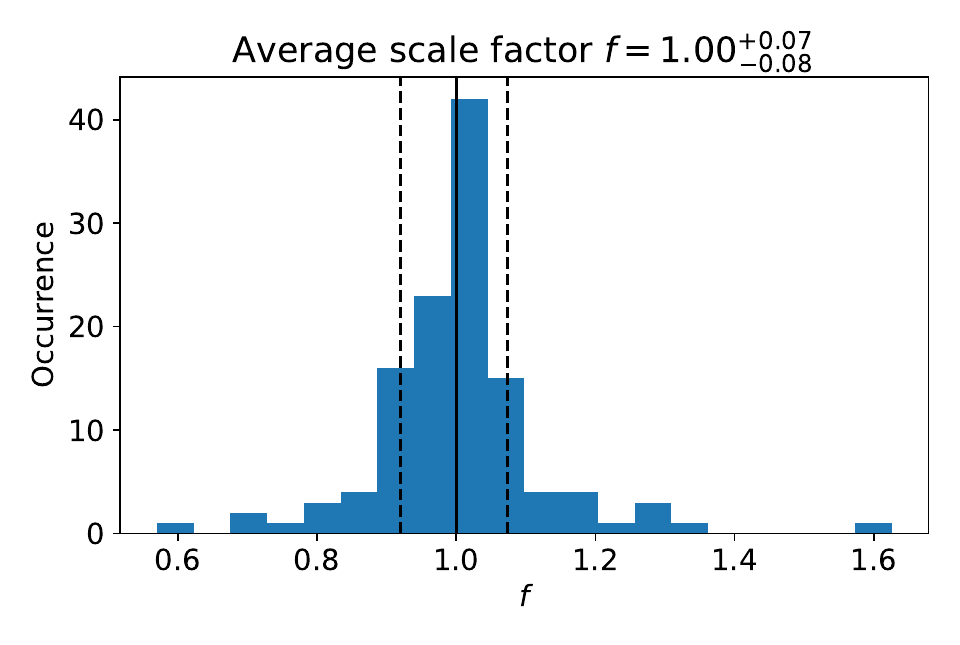}
    \caption{Distribution of scale factors applied to those GRAVITY companion spectra which were flux-calibrated using the host star. The spectra were scaled to bring individual epochs of the same object into agreement. The median scale factor is shown by the solid black line and the 16th and 84th percentiles are shown by the dashed black lines.}
    \label{fig:fs}
\end{figure}

\red{For scaling the individual GRAVITY epochs, we applied the following methodology.} Firstly, we determined the best fitting scale factors for all epochs where the host star was used to flux-calibrate the companion spectrum (on-star). This was done by minimizing the $\chi^2$ between each pair of spectra which is defined as
\begin{equation}
    \chi_{ij}^2 = (F_i - F_j)^T \cdot (\Sigma_i + \Sigma_j)^{-1} \cdot (F_i - F_j),
\end{equation}
while keeping the average scale factor equal to 1. \red{Here, $F_i$ are the companion flux spectra and $\Sigma_i$ are their covariances.} Then, we determined the best fitting scale factors for all epochs where a different star (e.g., the swap calibrator) was used to flux-calibrate the companion spectrum (off-star). This was done by minimizing the $\chi^2$ between a given off-star spectrum and all already scaled on-star spectra. If no on-star spectra were available for a given object, the off-star spectra were scaled in the same way as it was described above for the on-star spectra, that is, by minimizing the $\chi^2$ between each pair of spectra while keeping the average scale factor equal to 1. This procedure was chosen to give the flux calibration with the on-star reference a higher credibility than that with the off-star reference, because for the former the flux reference source was typically observed closer in time and in the same direction on the sky (i.e., at the same airmass) as the science target (the companion). We note that there are only two objects for which no on-star spectra were available, CD-35~2722~b and YSES~1~b. For these two objects, our absolute flux calibration should be taken with caution. The derived scale factors $f$ for all epochs are given in the last column of the observations table on Zenodo\footnote{Available at \url{https://doi.org/10.5281/zenodo.17295254}} and shown for all on-star observations in Figure~\ref{fig:fs}. For these on-star observations, the $1\sigma$ relative flux uncertainty is $\sim7$--8\%. An example of the above procedure for HD~206505~b is shown in Figure~\ref{fig:hd206505_b_all} (bottom panel).

\red{The scaled epochs for all 39 objects are shown in Figure~\ref{fig:all_spectra}. After the scaling, the different epochs appear consistent with each other to within their uncertainties. Therefore, we can now compute a combined (averaged) spectrum for each object as the covariance-weighted mean spectrum of all available epochs}
\begin{equation}
    F_\text{mean} = \Sigma_\text{mean} \cdot \sum_{i=1}^N \Sigma_i^{-1} \cdot F_i,
\end{equation}
where
\begin{equation}
    \Sigma_\text{mean} = \left(\sum_{i=1}^N \Sigma_i^{-1}\right)^{-1},
\end{equation}
$F_i$ are the companion flux spectra, $\Sigma_i$ are their covariances, and $N$ is the number of epochs available for a given object. \red{The final combined spectra for all 39 objects are also shown in Figure~\ref{fig:all_spectra} and we remind the reader that an additional systematic flux calibration uncertainty of $\pm10\%$ needs to be accounted for when working with these combined spectra. In agreement with this, we add a systematic flux calibration uncertainty -- a correlated error -- of 0.1~mag in quadrature to all objects in the remainder of this paper to account for the observed $\pm10\%$ systematic flux variations in the GRAVITY spectra.}

\subsection{Systematics in the GRAVITY spectra}
\label{sec:systematics_in_the_gravity_spectra}

The data reduction process relies on a dual calibration scheme. The first step involves calibrating the visibility measured by the science spectrometer using the visibility measured by the fringe tracker, obtained at exactly the same time. This corrects for losses in coherent flux caused by atmospheric turbulence and fringe tracking residuals, since both the science spectrometer and the fringe tracker experience the same perturbations simultaneously. The second calibration step involves bracketing the companion observations with observations of the host star -- or, in a few specific cases, a binary star system -- in order to derive the transfer function between the fringe tracker and the science spectrometer. This transfer function accounts for the relative transmission differences between the two interferometers. If this calibration is performed at the same airmass, it should also, in principle, correct for telluric absorption. This dual calibration approach explains how we are able to obtain high-quality companion spectra even within telluric water vapor absorption bands at the edges of the \textit{K}-band, including the strong absorption feature at $2.06~\micron$.

However, we have identified a limitation in this calibration scheme, which arises from the field of view of the single-mode fibers. These fibers have an extremely narrow field of view, equal to a single telescope’s diffraction limit. As a result, throughput drops significantly if the fiber is not precisely aligned with the companion (below 50\% at 35~mas pointing offset). Small pointing offsets can be partially corrected using an analytical estimate of the coupling losses \citep[see Appendix~A of][]{wang2021}. However, these corrections assume a perfect point spread function (PSF). If the PSF is altered due to optical aberrations, the theoretical correction no longer holds. Moreover, if the fiber is misaligned, the atmospheric impact on the coherent flux can differ significantly between the fringe tracker and the science spectrometer. This means that even the first calibration step can become biased. In fact, for several targets, we have observed discrepancies between spectra taken at different epochs that exceed the expected statistical uncertainties. In rare cases, calibration errors of up to $\sim60\%$ have been observed, adding a significant bias to the absolute flux calibration (see Figure~\ref{fig:fs}).

Beyond the issue of absolute flux calibration, additional wavelength-dependent spectral biases can arise from low-frequency fringing artifacts (commonly referred to as “wiggles”) that are not fully removed by polynomial fitting. This issue becomes more severe when the companion is located very close to its host star, where the coherent stellar flux is high. In these cases, a low-order polynomial is used to avoid subtracting the companion signal itself, which reduces the effectiveness of the stellar subtraction. Fortunately, such configurations are relatively rare (notable exceptions include HD~206893~c and 51~Eri~b). New data reduction techniques and observational strategies are currently under development to mitigate the impact of this parasitic fringing.

\subsection{Description of data format}
\label{sec:description_of_data_format}

The full ExoGRAVITY Spectral Library is available online\footnote{Available at \url{https://doi.org/10.5281/zenodo.17294469}}. For each companion, there is a FITS file with two extensions \red{containing the combined GRAVITY spectrum}. The first (primary) extension contains the combined companion flux spectrum. The three columns of data correspond to:
\begin{enumerate}
    \item the wavelength (in $\text{\textmu m}$),
    \item the flux (in $\text{W}/\text{m}^2/\text{\textmu m}$),
    \item the flux uncertainties (in $\text{W}/\text{m}^2/\text{\textmu m}$); \red{\textit{without}} the systematic flux calibration uncertainty of $10\%$.
\end{enumerate}
The second extension contains a binary table with the same information and an additional fourth column containing the covariance matrix of the companion spectrum. Furthermore, the primary header contains a header keyword for the name and the derived scale factor for each input epoch that was used to obtain the final combined spectrum. \red{There is also a similar FITS file with the GRAVITY flux spectrum from each individual epoch (file name contains observation date and fiber pointing).} The online library also contains a FITS file for each flux reference source (host star or swap calibrator). They contain the object's spectrum in native (to the model grid) spectral resolution in the same units as the companion FITS files. They also contain two additional extensions with the object's spectrum sampled down to a spectral resolution of $R \sim 500$ (GRAVITY medium) and $R \sim 4000$ (GRAVITY high). Furthermore, the primary header contains the inferred stellar parameters and the included archival photometry. We note that there are no covariances for the stellar model spectra since we propagate only the covariances from the GRAVITY contrast spectra.

\section{Results}
\label{sec:results}

The main result from this paper is the ExoGRAVITY Spectral Library${}^\text{2}$ of giant exoplanet and brown dwarf companions. This publicly available database contains science-ready \textit{K}-band GRAVITY spectra of 39 substellar objects and aims to make high-level science products from GRAVITY available to the wider community. We also conduct a basic atmospheric characterization of all 39 objects in the library based solely on their medium resolution ($R \sim 500$) GRAVITY spectra and an investigation of their evolutionary history in the following Sections. The two companions for which high resolution ($R \sim 4000$) GRAVITY spectra are available will be analyzed in separate publications by Ravet et al. in prep. (for $\beta$~Pic~b) and Kral et al. in prep. (for HD~206893~b).

\subsection{Spectral types}
\label{sec:spectral_types}

\begin{figure*}
    \centering
    \includegraphics[trim={0 0.5cm 0 0.5cm},clip,width=\textwidth]{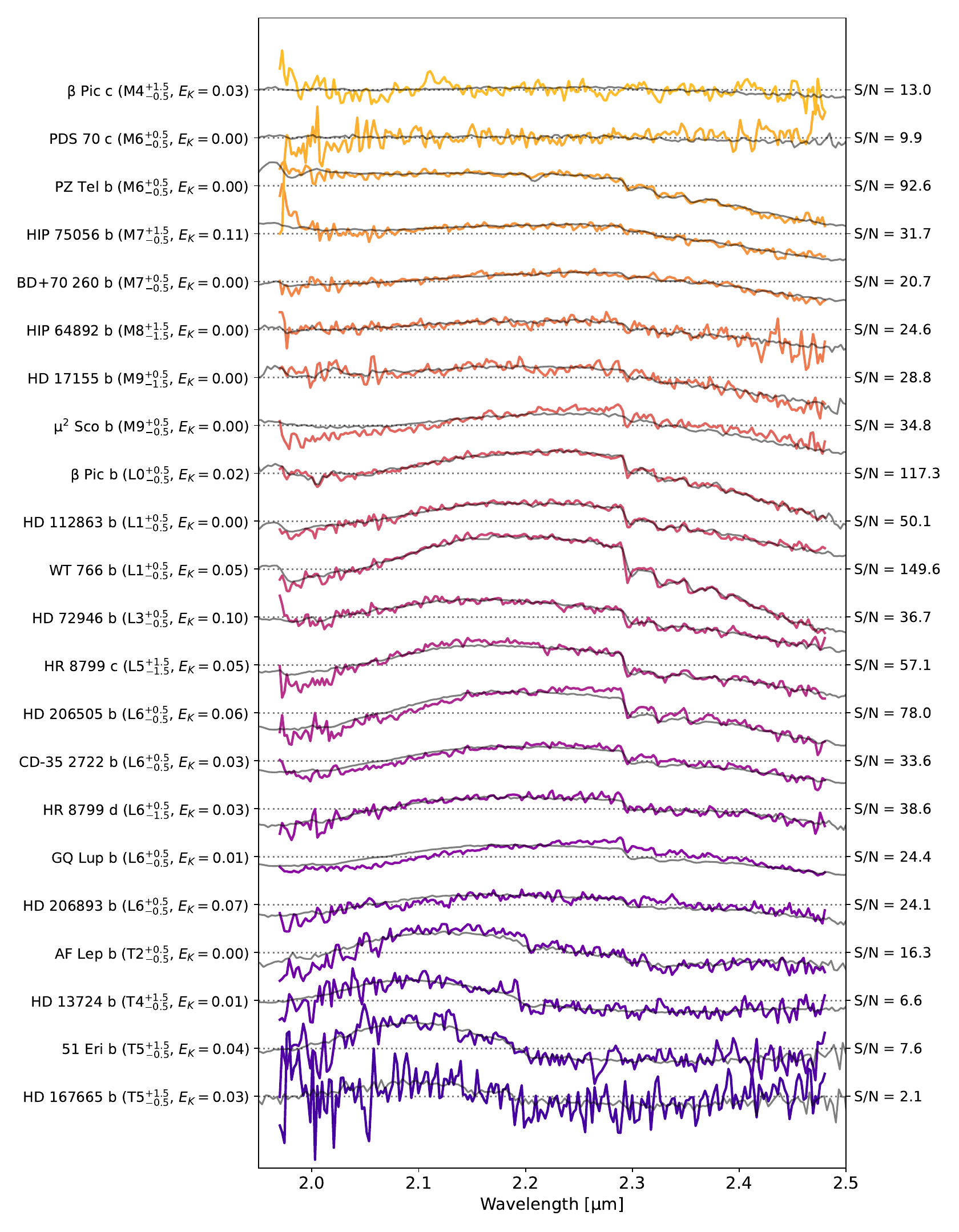}
    \caption{GRAVITY \textit{K}-band spectra at medium spectral resolution ($R \sim 500$) of the 22 substellar companions with well-constrained spectral type, sorted from M-type (top) to T-type (bottom). The spectra are scaled with respect to each other by the ratio of the square root of their SNR and their mean flux, and they are offset along the y-axis so that their mean value is located at the height of their corresponding tick mark for better readability. The best fitting spectral type templates from the SpeX Prism Libraries are shown in gray with their corresponding \textit{K1}-\textit{K2} extinction $E_K$ in units of magnitudes shown on the left. The SNR of each spectrum is shown on the right.}
    \label{fig:all_combined_spectra_good}
\end{figure*}

\begin{comment}
\begin{figure*}
    \centering
    \includegraphics[trim={0 0 0 0},clip,width=\textwidth]{figures/all_combined_spectra_poor_plain.pdf}
    \caption{Same as Figure~\ref{fig:all_combined_spectra_good}, but for the remaining 17 objects with poorly-constrained spectral type.}
    \label{fig:all_combined_spectra_poor}
\end{figure*}
\end{comment}

The near-infrared \textit{K}-band ($\sim2.0$--$2.4~\micron$) is a powerful spectral regime for the characterization of self-luminous giant planets and brown dwarfs \citep[e.g.,][]{oppenheimer1998,xuan2022}. It harbors absorption bands from CO and CH${}_4$ and its spectral slope provides clues on the nature of the object's atmosphere (clear vs. cloudy). To get a first idea of the 39 objects' physical properties, we determined their spectral types by matching their GRAVITY \textit{K}-band spectra with spectral type templates from the SpeX Prism Libraries \citep{burgasser2014}. The GRAVITY spectra analyzed here have a spectral resolution of $R \sim 500$ while the SpeX templates have a significantly lower spectral resolution \red{($R \sim 120$)}. Hence, we first binned down the GRAVITY spectra to the same spectral resolution as the SpeX templates. \red{We do not expect this binning to have any significant impact since the spectral types determined from the \textit{K}-band are mostly governed by the pseudo-continuum slope.}

For each template spectrum $k$ and each extinction value $A_V$ between 0--10~mag in 0.1~mag steps \citep[assuming the extinction law from][]{cardelli1989}, we computed the goodness-of-fit statistic $G_k$ between the template and the GRAVITY spectrum as defined in Equation~1 of \citet{cushing2008} using the atmospheric modeling toolkit \texttt{species}\footnote{Available at \url{https://github.com/tomasstolker/species}} \citep{stolker2020}. Then, we plotted the goodness-of-fit statistic as a function of the template spectral type considering only the extinction value yielding the smallest $\chi^2$ for each template. For some objects in the ExoGRAVITY Spectral Library with low SNR spectra, the fit would prefer an extinction $A_V > 10$~mag. In these cases, we did then fix the extinction at $A_V = 0$~mag, acknowledging that the SNR of the GRAVITY spectra is not sufficient to constrain the extinction. Given the large scatter in $G_k$ observed for individual spectral types, we also decided to only consider the minimum value of $G_k$ for each spectral type from here on. The best fitting spectral type was then taken as the spectral type with the smallest minimum $G_k$. Uncertainties for the spectral type were determined by computing the significance of the best fitting spectral type over all other spectral types according to a $\chi^2$-distribution, that is
\begin{equation}
    p = 1 - \text{CDF}_\nu\left(\frac{\nu\chi^2_\text{r,test}}{\chi^2_\text{r,best}}\right),
\end{equation}
where $\text{CDF}_\nu$ denotes the cumulative distribution function of a $\chi^2$-distribution with $\nu = N_\lambda$ degrees of freedom, $\chi^2_\text{r,test}$ is the reduced $\chi^2$ (here obtained as $G_k/N_\lambda$) of the other spectral type and $\chi^2_\text{r,best}$ is the reduced $\chi^2$ of the best fitting spectral type. The range of spectral types lying within $1\sigma$ of the best fitting spectral type was then adopted as the spectral type uncertainty. The above procedure is illustrated for all 39 companions in Figure~\ref{fig:all_spectral_types}. Most derived spectral types are consistent with those reported in the literature, for instance for 51~Eri~b \citep{macintosh2015}, $\beta$~Pic~b \citep{bonnefoy2014}, and HR~8799~c/d/e \citep{greenbaum2018}. However, for the PDS~70 protoplanets \citep{mesa2019} and the close-in $\beta$~Pic~c and HD~206893~c companions our spectral types are systematically off. We attribute this to the low SNR resulting in apparently flat spectra, or instrumental noise at short separations (see Section~\ref{sec:systematics_in_the_gravity_spectra}).

\subsection{GRAVITY \textit{K}-band spectra}
\label{sec:gravity_k-band_spectra}

\begin{figure}
    \centering
    \includegraphics[trim={0 0 0 0},clip,width=\columnwidth]{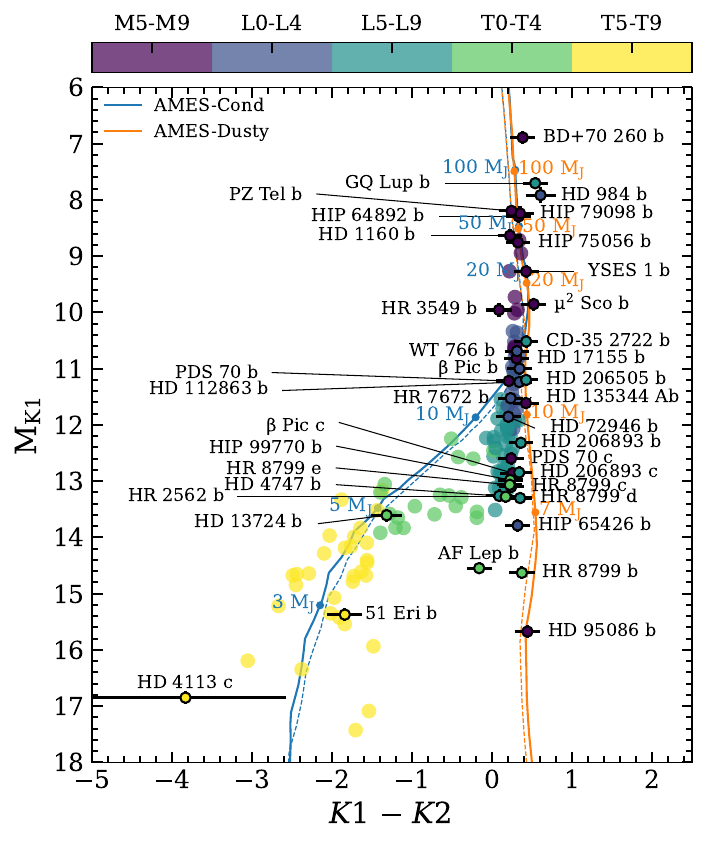}
    \caption{\textit{K}1-\textit{K}2 color magnitude diagram for 38 of the 39 objects in the ExoGRAVITY Spectral Library. AMES-Cond and AMES-Dusty evolutionary tracks at an age of 20~Myr (solid lines) and 100~Myr (dashed lines) are plotted in the background and mid-M- to late-T-type objects from the SpeX Prism Libraries are shown for reference. HD~167665~b is missing from the plot because its methane absorption is strong yielding only an upper flux limit in \textit{K}2.}
    \label{fig:color_magnitude}
\end{figure}

\begin{figure*}
    \centering
    \includegraphics[trim={0 0 0 0},clip,width=0.9\textwidth]{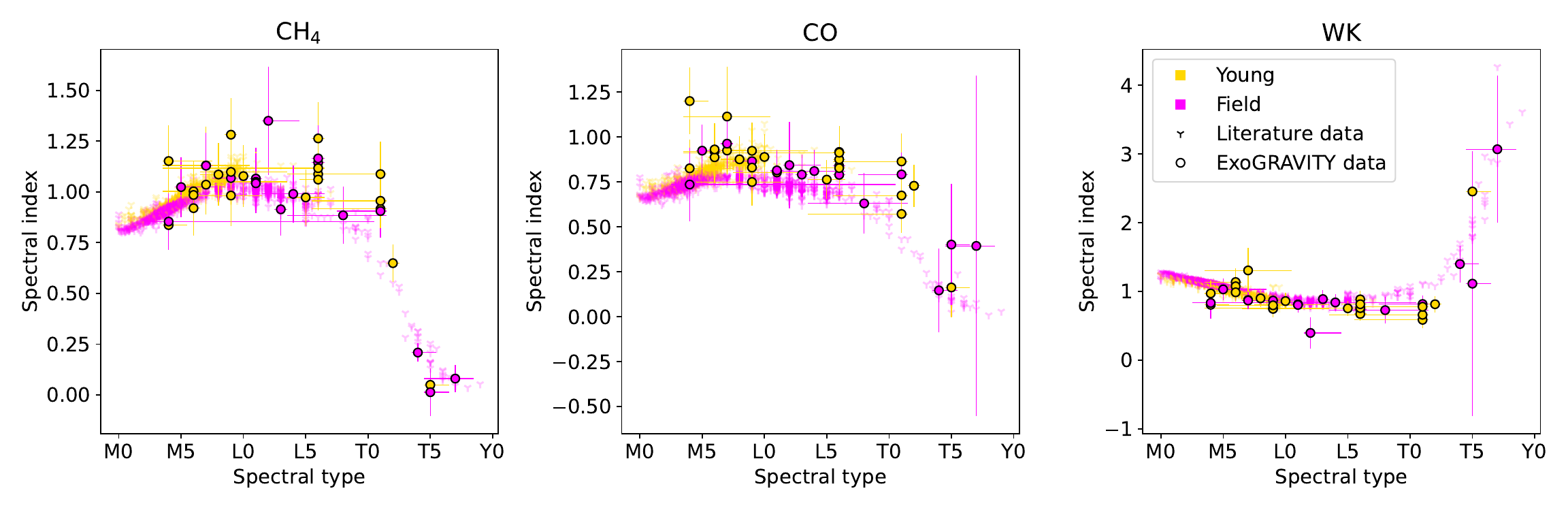}
     \caption{Methane- (CH${}_4$), carbon-monoxide- (CO), and water- (WK) sensitive spectral indices for all objects in the ExoGRAVITY Spectral Library. For reference, young ($<10$~Myr) and field dwarfs from the literature are shown in the background. For the ExoGRAVITY sample, objects are considered young if they have an estimated age of $<50$~Myr or if they are members of the Argus moving group.}
     \label{fig:spectral_indices}
\end{figure*}

\red{Starting with low-mass M-dwarfs with rather featureless spectra in the \textit{K}-band (at a spectral resolution of $R \sim 500$) at the top, over cloudy L-dwarfs with increasingly strong CO absorption bands at $\sim2.3~\micron$ in the middle, to methane-dominated T-dwarfs with broad absorption features removing most of the flux beyond $\sim2.2~\micron$ at the bottom, Figure~\ref{fig:all_combined_spectra_good} shows the GRAVITY \textit{K}-band spectra of all companions with a well-constrained spectral type. This is here defined as having a spectral type uncertainty of less than two subtypes which captures all those objects whose visual spectral features agree well with the features expected for their spectral type. A similar figure for the 17 objects with poorly-constrained spectral type can be found on Zenodo\footnote{Available at \url{https://doi.org/10.5281/zenodo.17295254}}.} The SNR of each spectrum is computed as the 90th percentile of the ratio of the flux and the flux uncertainties. Therefore, the resulting SNR values are reached over at least a 10\% bandpass within the GRAVITY \textit{K}-band and reflect an outlier-safe estimation of the maximum achieved SNR. The choice to focus the SNR estimation on a narrow 10\% bandpass was made because especially for the T-dwarfs, more than 50\% of the GRAVITY bandpass can contain zero flux due to the strong methane absorption longward of $\sim2.2~\micron$.

Figure~\ref{fig:all_combined_spectra_good} illustrates the \textit{K}-band spectral signatures of the substellar L-T sequence. This is because we only plotted those spectra with a well-constrained spectral type. The spectra of the remaining 17 companions with poorly-constrained spectral type have lower SNR as spectra with a similar but well-constrained spectral type and are sometimes affected by residual systematic noise manifesting itself as ``wiggles'' in the spectra (see also Section~\ref{sec:systematics_in_the_gravity_spectra}).

The near-infrared \textit{K}-band is a powerful spectral regime to study the atmospheric carbon chemistry of substellar companions \citep[e.g.,][]{kirkpatrick2005}. At the spectral resolution of GRAVITY, CO absorption bands can be seen at $\sim2.3~\micron$ for the hotter L-dwarfs. For the cooler T-dwarfs, the CO $\leftrightarrow$ CH${}_4$ chemical equilibrium shifts towards methane and a broad absorption trough longward of $\sim2.2~\micron$ appears in the GRAVITY spectra removing almost the entire flux at these wavelengths. Surface gravity-dependent effects like vertical mixing are expected to change the effective temperature at which substellar companions transition from CO-rich to CH${}_4$-rich photospheres and from red to blue near-infrared colors \citep[e.g.,][]{zahnle2014}. To investigate this transition for our sample of 39 substellar companions, we computed synthetic SPHERE \textit{K}1- and \textit{K}2-band magnitudes \red{\citep{dohlen2008}} from the GRAVITY \textit{K}-band spectra for each object and displayed them in a color-magnitude diagram (Figure~\ref{fig:color_magnitude}). It can be seen that most objects sit along the AMES evolutionary track for dusty atmospheres. Only a handful of companions appear to have clear atmospheres (HD~4113~c, 51~Eri~b, and HD~13724~b). In addition, AF~Lep~b seems to be located somewhere along the substellar L-T transition where the atmospheres are changing from cloudy to clear. \red{The color-magnitude diagram also shows that for massive objects such as HD~13724 \citep[dynamical mass of $\sim36$--$50~\text{M}_\text{Jup}$,][]{rickman2019,rickman2020,brandt2021} and the field dwarfs from the SpeX Prism Libraries the substellar L-T transition from cloudy to clear atmospheres occurs earlier (i.e., at brighter absolute magnitude and thus higher effective temperature) than for less massive objects such as HR~8799~b \citep[$5.8\pm0.4~\text{M}_\text{Jup}$,][]{zurlo2022} and AF~Lep~b \citep[$\sim2$--$7~\text{M}_\text{Jup}$,][]{derosa2023,franson2023,mesa2023}.} This observed behavior can provide clues about the nature of the physical processes underlying the substellar L-T transition which are thought to be dependent on the surface gravity and potentially also the metallicity of the object \citep[e.g.,][]{burrows2006}.

\subsection{Spectral indices}
\label{sec:spectral_indices}

Spectral indices have been used in the literature to determine spectral types or to assess the youth or surface gravity of substellar objects. This has been possible because the spectral characteristics of M-, L-, and T-type dwarfs have been observed to vary strongly with surface gravity \citep[e.g.,][]{lucas2001,gorlova2003,allers2013}. Spectral indices are usually defined as the ratio of the flux over two distinct bandpasses which often probe the depth of a certain molecular feature compared to the continuum. We identified three spectral indices that are probing different chemical species in the GRAVITY \textit{K}-band and computed them for all objects in the ExoGRAVITY Spectral Library. These are the CH${}_4$-C \citep{burgasser2002}, CO \citep{mclean2003}, and WK \citep{weights2009} indices from Table~D.1 of \citet{piscarreta2024}. Figure~\ref{fig:spectral_indices} shows these three indices plotted against the spectral types that we determined in Section~\ref{sec:spectral_types}. The literature samples for young and for field dwarfs compiled in \citet{piscarreta2024} are also shown for reference. They consist of objects down to L5 from \citet{almendros-abad2022} combined with objects later than L5 from the SpeX Prism Libraries \citep{burgasser2014} and the Montreal Spectral Library \citep{gagne2015}. This enables us to probe whether there is agreement between the literature samples and the substellar companions in the ExoGRAVITY Spectral Library. \red{Here, objects are considered young if they have an estimated age of $<10$~Myr for the literature samples \citep[in agreement with the analysis presented in][]{piscarreta2024} and of $<50$~Myr for the ExoGRAVITY sample\footnote{A larger age range of 50~Myr is considered for the ExoGRAVITY sample because there are only three objects with an age of $<10$~Myr which is not a sufficient sample size to look for population scale trends between younger and older objects.}.} Furthermore, we consider members of the Argus moving group as young since the age of this moving group varies widely in the literature, but might be as young as 40--50~Myr \citep{zuckerman2019}.

Figure~\ref{fig:spectral_indices} reveals that for all three shown spectral indices, the ExoGRAVITY companions broadly follow the evolution of the literature samples. Both the CH${}_4$- and the CO-sensitive indices show a strong change in spectral index across the substellar L-T transition, where a shift from CO- to CH${}_4$-dominated photospheres can be observed. For the literature samples, there is also a bifurcation between young and field objects with young objects typically having slightly elevated spectral indices. This makes these two spectral indices useful for youth or gravity assessment. A similar trend with age can be seen in the ExoGRAVITY sample, although the uncertainties are too large to re-detect this bifurcation in the latter alone. The water-sensitive WK index shows a significant gradient on both sides of the L-T transition, but with an opposite sign. It also shows no significant bifurcation as a function of age. Therefore, it is well suited to determine the spectral type of substellar objects directly from the value of the spectral index itself, as long as it is clear on which side of the L-T transition a give object lies. While the literature data shown here consists mostly of isolated field dwarfs, our ExoGRAVITY sample consists of bound substellar companions. Nevertheless, the ExoGRAVITY sample agrees with the literature data within the uncertainties and we conclude that established spectral indices for field dwarfs can also be used for characterizing bound substellar companions.

\section{Discussion}
\label{sec:discussion}

\begin{table*}
\caption{\red{Derived companion parameters and stellar association membership from the literature.}}
\label{tab:companion_parameters}
\centering
\begin{tabular}{c c c c c c c c c c c c c c c c}
\hline\hline
Object & Member & Age [Myr] & $\log L/\text{L}_\odot$ [dex] & $\log L_\text{G}/\text{L}_\odot$ [dex] & $\Delta$ [$\sigma_\text{G}$] & $M_\text{dyn}$ [$\text{M}_\text{Jup}$] & $M_\text{evo}$ [$\text{M}_\text{Jup}$] & Refs.\\
\hline
BD+70 260 b & FIELD & -- & -- & $-2.00^{+0.10}_{-0.10}$ & -- & -- & -- & 1\\
51 Eri b & BPMG & $24^{+3}_{-3}$ & $-5.48^{+0.13}_{-0.13}$ & $-5.21^{+0.63}_{-0.27}$ & 1.0 & -- & $4.6^{+3.3}_{-1.7}$ & 1,2,3\\
AF Lep b & BPMG & $24^{+3}_{-3}$ & $-5.2^{+0.1}_{-0.2}$ & $-5.15^{+0.11}_{-0.14}$ & 0.3 & $3.75^{+0.5}_{-0.5}$ & $4.0^{+0.5}_{-0.5}$ & 1,3,4\\
$\beta$ Pic b & BPMG & $24^{+3}_{-3}$ & $-3.78^{+0.03}_{-0.03}$ & $-3.73^{+0.07}_{-0.14}$ & 0.4 & $11.90^{+2.93}_{-3.04}$ & $13.5^{+0.5}_{-0.5}$ & 1,3,5,6\\
$\beta$ Pic c & BPMG & $24^{+3}_{-3}$ & $-4.50^{+0.04}_{-0.04}$ & $-4.45^{+0.13}_{-0.09}$ & 0.6 & $8.89^{+0.75}_{-0.75}$ & $8.7^{+1.5}_{-1.4}$ & 1,3,5,6\\
CD-35 2722 b & ABDMG & $149^{+51}_{-19}$ & -- & $-3.80^{+0.08}_{-0.13}$ & -- & -- & $39.2^{+4.3}_{-5.6}$ & 1,3\\
GQ Lup b & UCL & $3.5^{+1.5}_{-1.5}$ & $-2.16$ & $-2.37^{+0.08}_{-0.12}$ & 2.6 & -- & $26.4^{+2.9}_{-3.8}$ & 1,7,8\\
HD 984 b & FIELD & $115^{+85}_{-85}$ & $-2.88^{+0.03}_{-0.03}$ & $-2.51^{+0.28}_{-0.57}$ & 0.7 & $61^{+4}_{-4}$ & $185^{+124}_{-82}$ & 1,5\\
HD 1160 b & FIELD & $100^{+200}_{-70}$ & $-2.99^{+0.02}_{-0.03}$ & $-2.74^{+0.14}_{-0.11}$ & 2.2 & -- & $103^{+25}_{-22}$ & 1,9,10\\
HD 4113 c & FIELD & $5000^{+1300}_{-1700}$ & -- & $-5.91^{+0.90}_{-0.39}$ & -- & $66^{+5}_{-4}$ & $33.8^{+18.8}_{-13.6}$ & 1,11\\
HD 4747 b & FIELD & $3600^{+600}_{-600}$ & $-4.54^{+0.06}_{-0.06}$ & $-4.62^{+0.19}_{-0.20}$ & 0.4 & $66.2^{+2.5}_{-2.5}$ & $64.0^{+5.8}_{-7.0}$ & 1,12\\
HD 13724 b & FIELD & $2800^{+500}_{-500}$ & $-4.78^{+0.07}_{-0.07}$ & $-4.60^{+0.34}_{-0.29}$ & 0.6 & $36.2^{+1.6}_{-1.6}$ & $60.8^{+10.0}_{-10.8}$ & 1,13\\
HD 17155 b & FIELD & -- & -- & $-3.57^{+0.10}_{-0.10}$ & -- & $86.0^{+0.4}_{-0.5}$ & -- & 1,14\\
HD 72946 b & FIELD & $2670^{+250}_{-490}$ & $-4.14^{+0.01}_{-0.01}$ & $-4.04^{+0.07}_{-0.09}$ & 1.1 & $69.5^{+0.5}_{-0.5}$ & $77.4^{+1.3}_{-1.2}$ & 1,15\\
HD 95086 b & LCC & $27^{+3}_{-3}$ & $-5.00^{+0.03}_{-0.05}$ & $-4.93^{+0.15}_{-0.14}$ & 0.5 & -- & $5.2^{+0.9}_{-0.7}$ & 1,16\\
HD 112863 b & FIELD & $3310^{+2910}_{-2910}$ & $-3.79^{+0.03}_{-0.03}$ & $-3.78^{+0.07}_{-0.09}$ & 0.0 & $77.1^{+2.9}_{-2.8}$ & -- & 1,17,18\\
HD 167665 b & FIELD & $6200^{+1130}_{-1130}$ & $-4.89^{+0.02}_{-0.03}$ & $-4.89^{+0.44}_{-0.69}$ & 0.0 & $60.3^{+0.7}_{-0.7}$ & $60.1^{+11.3}_{-15.5}$ & 1,19\\
HD 206505 b & FIELD & $3940^{+2510}_{-2510}$ & $-3.71^{+0.02}_{-0.02}$ & $-3.79^{+0.07}_{-0.14}$ & 1.2 & $79.8^{+1.8}_{-1.8}$ & -- & 1,17,18\\
HD 206893 b & ARG & $155^{+15}_{-15}$ & $-4.23^{+0.01}_{-0.01}$ & $-4.24^{+0.10}_{-0.09}$ & 0.1 & $28.0^{+2.2}_{-2.1}$ & $28.1^{+2.9}_{-3.3}$ & 1,20\\
HD 206893 c & ARG & $155^{+15}_{-15}$ & $-4.42^{+0.02}_{-0.01}$ & $-4.42^{+0.12}_{-0.10}$ & 0.0 & $12.7^{+1.2}_{-1.0}$ & $21.7^{+4.3}_{-7.8}$ & 1,20\\
HIP 64892 b & LCC & $15^{+3}_{-3}$ & $-2.66^{+0.10}_{-0.10}$ & $-2.57^{+0.11}_{-0.11}$ & 0.8 & -- & $40.5^{+10.2}_{-11.7}$ & 1,21,22\\
HIP 65426 b & LCC & $15^{+3}_{-3}$ & $-4.15^{+0.01}_{-0.01}$ & $-4.06^{+0.10}_{-0.09}$ & 1.0 & -- & $10.8^{+0.9}_{-1.2}$ & 1,21,23\\
HIP 75056 b & UCL & $16^{+2}_{-2}$ & $-2.8^{+0.3}_{-0.3}$ & $-2.72^{+0.10}_{-0.12}$ & 0.7 & $35^{+10}_{-10}$ & $30.8^{+9.2}_{-8.0}$ & 1,21,24\\
HIP 79098 b & USCO & $10^{+3}_{-3}$ & -- & $-2.66^{+0.21}_{-0.33}$ & -- & -- & $33.8^{+23.2}_{-12.2}$ & 1,21\\
HIP 99770 b & ARG & $220^{+180}_{-180}$ & $-4.52^{+0.01}_{-0.01}$ & $-4.30^{+0.16}_{-0.15}$ & 1.5 & $17^{+6}_{-5}$ & $33.3^{+10.6}_{-12.2}$ & 1,25,26\\
HR 2562 b & FIELD & $475^{+275}_{-275}$ & -- & $-4.63^{+0.10}_{-0.11}$ & -- & -- & $33.6^{+8.8}_{-9.1}$ & 1,27\\
HR 3549 b & FIELD & $125^{+25}_{-25}$ & -- & $-3.33^{+0.18}_{-0.22}$ & -- & -- & $52.0^{+14.2}_{-10.0}$ & 1,28\\
HR 7672 b & FIELD & $1920^{+300}_{-300}$ & $-4.19^{+0.04}_{-0.04}$ & $-3.92^{+0.09}_{-0.10}$ & 2.8 & $72.7^{+0.8}_{-0.8}$ & $78.6^{+1.8}_{-2.0}$ & 1,12\\
HR 8799 b & COL & $42^{+24}_{-24}$ & $-5.21^{+0.07}_{-0.07}$ & $-5.09^{+0.12}_{-0.12}$ & 1.0 & -- & $6.1^{+1.6}_{-1.5}$ & 1,13,29\\
HR 8799 c & COL & $42^{+24}_{-24}$ & $-4.69^{+0.04}_{-0.03}$ & $-4.58^{+0.08}_{-0.08}$ & 1.4 & -- & $9.4^{+1.4}_{-2.4}$ & 1,13,29\\
HR 8799 d & COL & $42^{+24}_{-24}$ & $-4.62^{+0.03}_{-0.03}$ & $-4.66^{+0.08}_{-0.11}$ & 0.4 & -- & $8.7^{+1.7}_{-2.2}$ & 1,13,29\\
HR 8799 e & COL & $42^{+24}_{-24}$ & $-4.74^{+0.04}_{-0.04}$ & $-4.53^{+0.08}_{-0.10}$ & 2.2 & $9.6^{+1.9}_{-1.9}$ & $9.8^{+1.5}_{-2.1}$ & 1,13,29\\
$\mu^2$ Sco b & UCL & $16^{+2}_{-2}$ & -- & $-3.17^{+0.08}_{-0.09}$ & -- & -- & $19.8^{+3.8}_{-4.0}$ & 1,21\\
PDS 70 b & UCL & $5.4^{+1.0}_{-1.0}$ & $-3.76^{+0.22}_{-0.22}$ & $-3.80^{+0.14}_{-0.13}$ & 0.3 & -- & $8.8^{+1.3}_{-1.3}$ & 1,30,31\\
PDS 70 c & UCL & $5.4^{+1.0}_{-1.0}$ & $-4.09^{+0.39}_{-0.38}$ & $-4.33^{+0.14}_{-0.11}$ & 1.7 & -- & $5.0^{+0.9}_{-0.7}$ & 1,30,31\\
PZ Tel b & BPMG & $24^{+3}_{-3}$ & -- & $-2.46^{+0.07}_{-0.07}$ & -- & -- & $64.3^{+8.3}_{-7.9}$ & 1,3\\
HD 135344 Ab & UCL & $16^{+2}_{-2}$ & $-3.9^{+0.1}_{-0.1}$ & $-3.87^{+0.14}_{-0.14}$ & 0.2 & -- & $12.2^{+0.8}_{-1.1}$ & 1,21,32\\
WT 766 b & FIELD & $670^{+40}_{-40}$ & -- & $-3.58^{+0.06}_{-0.08}$ & -- & $77.3^{+1.4}_{-1.3}$ & -- & 1,33\\
YSES 1 b & LCC & $15^{+3}_{-3}$ & $-3.17^{+0.05}_{-0.05}$ & $-3.20^{+0.12}_{-0.11}$ & 0.3 & -- & $19.2^{+5.3}_{-3.7}$ & 1,21,34\\
\hline
\end{tabular}
\tablefoot{\red{$L$ is the luminosity from the literature and $L_\text{G}$ is the luminosity derived from the GRAVITY spectra. $\Delta$ is their difference in units of standard deviations of the GRAVITY-derived luminosity $\sigma_\text{G}$. An additional systematic uncertainty of 10\% ($\sim0.05$~dex) has been added in quadrature to the GRAVITY-derived luminosity to account for flux calibration errors. $M_\text{dyn}$ is the dynamical mass from the literature and $M_\text{evo}$ is the mass derived from evolutionary models based on the GRAVITY luminosity.} (1) \citealt{gagne2018}, (2) \citealt{brown-sevilla2023}, (3) \citealt{bell2015}, (4) \citealt{balmer2025}, (5) \citealt{franson2022}, (6) \citealt{lacour2021}, (7) \citealt{donati2012}, (8) \citealt{stolker2021}, (9) \citealt{maire2016}, (10) \citealt{sutlieff2024}, (11) \citealt{cheetham2018}, (12) \citealt{brandt2019}, (13) \citealt{brandt2021}, (14) \citealt{unger2023}, (15) \citealt{balmer2023}, (16) \citealt{malin2024}, (17) \citealt{rickman2024}, (18) \citealt{ceva2025}, (19) \citealt{maire2024}, (20) \citealt{hinkley2023}, (21) \citealt{pecaut2016}, (22) \citealt{cheetham2018b}, (23) \citealt{carter2023}, (24) \citealt{balmer2024}, (25) \citealt{currie2023}, (26) \cite{winterhalder2025}, (27) \citealt{mesa2018}, (28) \citealt{mesa2016}, (29) \citealt{nasedkin2024}, (30) \citealt{mueller2018}, (31) \citealt{wang2021}, (32) \citealt{stolker2025}, (33) \citealt{winterhalder2024}, (34) \citealt{bohn2020}.}
\end{table*}

\begin{figure*}
    \centering
    \includegraphics[trim={0 0 0 0},clip,width=0.9\columnwidth]{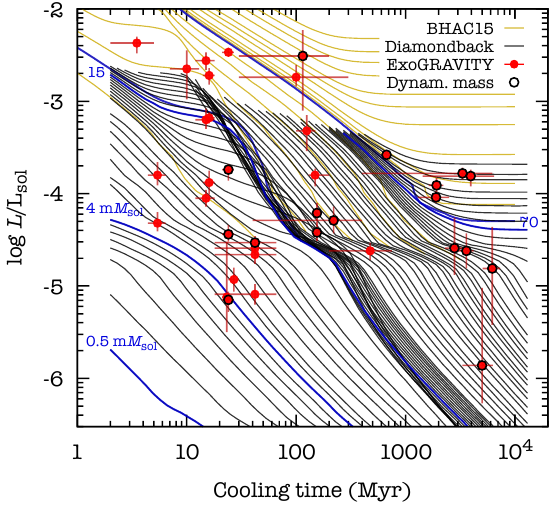}~~~~
    \includegraphics[trim={0 0 0 0},clip,width=0.9\columnwidth]{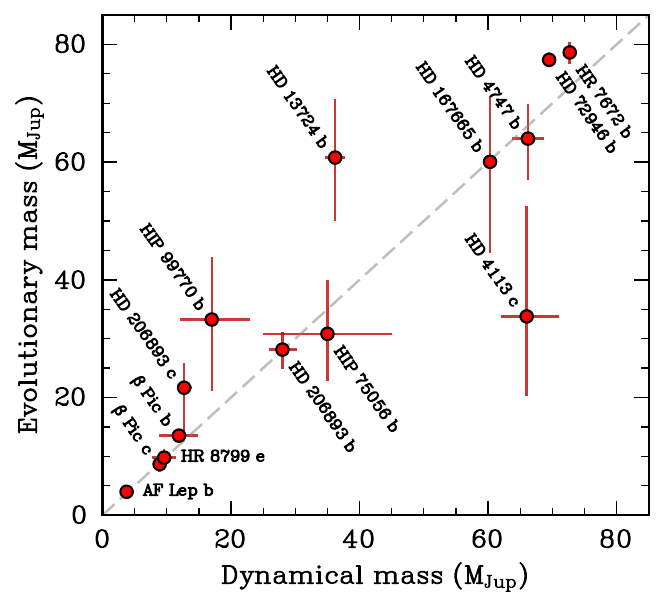}
    \caption{\textit{Left panel}: Sonora Diamondback hybrid gravity solar metallicity (\citealp{morley2024}) and BHAC15 (\citealp{baraffe2015}) cooling curves for substellar companions. The 37 objects from the ExoGRAVITY Spectral Library with age constraints available in the literature are shown by red points, and those with dynamical mass measurements have a black contour. Cooling curves for a few masses are highlighted in blue (1~m$\text{M}_\text{sol} = 1.048~\text{M}_\text{Jup}$). \textit{Right panel}: Correlation between dynamical mass constraints from the literature and evolutionary mass constraints from GRAVITY. HD~984~b lies outside the shown mass range but is still consistent between its evolutionary and dynamical mass to within $2\sigma$.}
    \label{fig:evolutionary_models}
\end{figure*}

\subsection{Atmospheric characterization}
\label{sec:atmospheric_characterization}

To derive bulk properties for the 39 objects in the ExoGRAVITY Spectral Library, we fitted atmospheric models to their GRAVITY spectra and investigated how well the \textit{K}-band constrains the luminosity, effective temperature, and radius. For this exercise, we considered the AMES-Cond and AMES-Dusty grids \citep{chabrier2000}, the BT-Cond, BT-Dusty, and BT-Settl grids \citep{allard2012}, the DRIFT-PHOENIX grid \citep{helling2008}, and the Exo-REM grid \citep{baudino2015,charnay2018}. All seven grids were fitted to the GRAVITY spectra of all 39 objects. Then, the objects were grouped according to which grids provided a good fit and which ones did not based on the reduced $\chi^2$ metric. The fitting was done with \texttt{species} \citep{stolker2020} which employs nested sampling with \texttt{PyMultiNest} \citep{buchner2014} to infer the model parameter posterior distributions. The T- and late L-type objects were found to be better fit by clear and hybrid grids (AMES-Cond, BT-Cond, BT-Settl, Exo-REM) while the early L- and M-type objects were found to be better described by cloudy and hybrid grids (AMES-Dusty, BT-Dusty, BT-Settl, DRIFT-PHOENIX, Exo-REM). Seven objects did not seem to show a preference for any of the explored grids because their spectra are dominated by noise. We note that these seven objects also have poorly-constrained spectral types according to the analysis in Section~\ref{sec:spectral_types}.

Figure~\ref{fig:all_lum_posteriors} shows the posterior distributions of the bolometric luminosity. Similar figures for the effective temperature and the radius can be found on Zenodo\footnote{Available at \url{https://doi.org/10.5281/zenodo.17295254}}. The typical $1\sigma$ uncertainty on the inferred bolometric luminosity is $\sim0.1$~dex. The precision of the luminosity constraint naturally correlates with the SNR of the GRAVITY spectrum from which it was derived. The luminosity measured from the GRAVITY \textit{K}-band spectra alone is always within $3\sigma$ of literature values where available and hence in statistical agreement with them. In most cases, these literature values were derived from data with a broader wavelength coverage than the GRAVITY spectra and are hence more reliable. Before the luminosities obtained from the GRAVITY spectra can be used to constrain the evolutionary state of the 39 objects, an additional systematic uncertainty of 10\% ($\sim0.05$~dex) must be added in quadrature due to the systematic uncertainties in the GRAVITY flux calibration (see Section~\ref{sec:combining_spectra_from_multiple_epochs}). The inferred effective temperature of the 39 objects is usually rather poorly constrained based on the GRAVITY \textit{K}-band spectra alone. Uncertainties of several hundreds up to a thousand Kelvin are the norm. For an individual object, the posteriors from different atmospheric model grids can be discrepant due to correlations between the effective temperature, the radius, and the assumed cloud prescription which cannot be resolved with the limited wavelength coverage of the GRAVITY spectra alone. Hence, the uncertainty in the derived effective temperature is typically dominated by systematic differences between the various atmospheric model grids. Individual models yield tighter constraints for apparently brighter objects whose spectra have higher SNR. The same is the case for the inferred radius of the 39 objects.

\subsection{Comparison with evolutionary models}
\label{sec:comparison_with_evolutionary_models}

Given the bolometric luminosity constraints from the GRAVITY spectra from Section~\ref{sec:atmospheric_characterization}, we then attempted to derive masses for all 39 objects using evolutionary models for substellar companions. This required an age constraint from young moving group membership or stellar age indicators for instance. We hence compiled age constraints from the literature for 37 of the 39 objects in Table~\ref{tab:companion_parameters}. Unfortunately, there was no age constraint available for two objects (BD+70~260~b and HD~17155~b). Moreover, the age of three objects (HD~112863~b, HD~206505~b, and WT~766~b) was derived from the dynamical mass of the companion using evolutionary models. Therefore, we could not use those ages for constraining the mass of the companions using evolutionary models as this would lead to a circular argument. For the other 34 objects, we used Sonora Diamondback hybrid gravity solar metallicity evolutionary models \citep{morley2024} to derive mass constraints. Seven objects were outside the luminosity range covered by these models so that we used the models from \citet{baraffe2015} instead (see Figure~\ref{fig:evolutionary_models}, left panel). The posterior distributions of the evolutionary model parameters (age, mass, effective temperature, radius) were inferred using \texttt{species} \citep{stolker2020}. The age constraints from the literature were used as Gaussian age priors with symmetric uncertainties taken as the average of the asymmetric error bars from the literature.

Table~\ref{tab:companion_parameters} shows the derived mass constraints from our evolutionary model analysis. For HD~206893~c, HIP~99770~b, $\mu^2$~Sco~b, and YSES~1~b we find bimodal mass posteriors with one peak below $\sim15~\text{M}_\text{Jup}$ and another one above $\sim15~\text{M}_\text{Jup}$ depending on whether the object started deuterium burning in its core or not \citep[e.g.,][]{hinkley2023}. In all but two cases, the evolutionary and dynamical masses are consistent with one another to well within $2\sigma$ (see Figure~\ref{fig:evolutionary_models}, right panel). The only two outliers are HD~72946~b and HR~7672~b for which we find slightly higher evolutionary masses if compared to their dynamical masses. This might point to issues with the evolutionary models, the dynamical masses, or the age constraints of these objects. Besides, for HD~4113~c, a brown dwarf companion at $\sim700$~mas separation from its host star, we can confirm the underluminosity already reported by \citet{cheetham2018}. We note that the evolutionary mass derived from the GRAVITY luminosity has rather large uncertainties since the luminosity itself remains poorly constrained from the atmospheric model fits (see Figure~\ref{fig:all_lum_posteriors}). When focusing on the BT-Cond grid alone, akin to the analysis in \citet{cheetham2018}, the GRAVITY-derived luminosity and therefore the evolutionary mass would be even lower than what is shown in Figure~\ref{fig:evolutionary_models}. Such a discrepancy between the dynamical and evolutionary mass could point to the existence of an unseen companion and HD~4113~c being a binary brown dwarf, as it was shown recently in the case of Gl~229~B \citep{xuan2024,whitebook2024}. Finally, we find an evolutionary mass of $185^{+124}_{-82}~\text{M}_\text{Jup}$ for HD~984~b which has a dynamical mass constraint of $61^{+4}_{-4}~\text{M}_\text{Jup}$ from \citet{franson2022}. We note that the two are consistent within $\sim1.5\sigma$. The large uncertainties of the evolutionary mass derived from the GRAVITY luminosity are not surprising given that the GRAVITY spectrum of HD~984~b is one of the spectra with the lowest SNR in the ExoGRAVITY Spectral Library.

\section{Summary and conclusions}
\label{sec:summary_and_conclusions}

We have compiled the ExoGRAVITY Spectral Library, a homogeneous library of 39 giant planet and brown dwarf companion \textit{K}-band spectra observed with VLTI/GRAVITY. This uniformly processed library aims to aid the identification of empirical trends among larger samples of substellar companions and to provide insight into the governing atmospheric processes on both sides of the L-T transition. Another recent example of a spectral library for VLT/SINFONI spectra of substellar objects at the M-L transition was published by \citet{palmabifani2025}. However, compared to SINFONI and other current and past instruments, the superior starlight suppression capabilities of GRAVITY, thanks to its usage of long-baseline interferometry, open a unique window for $R \sim 500$ spectroscopic characterization of exoplanets and brown dwarf companions close to their host star, down to a few au separations.

For most of the 39 objects in the ExoGRAVITY Spectral Library, multiple epochs of data are available. We observe flux variations between these individual epochs on the $\pm10\%$ level which we attribute to mostly instrumental systematics. Given the large amounts of data analyzed in this paper ($>150$ individual GRAVITY epochs amounting to $>100$ hours of on-source integration time), we are now able to robustly characterize the stability and uncertainties of our flux calibration scheme (see Section~\ref{sec:combining_spectra_from_multiple_epochs}). If the systematics in the GRAVITY spectra can be better understood and calibrated in the future (for instance by exploiting the data collected by the GRAVITY acquisition camera), the ExoGRAVITY Spectral Library might become a valuable resource for studying giant planet and brown dwarf variability. However, at the time of writing, the instrumental systematics still dominate any expected astrophysical variability by a factor of 5--10.

We derive spectral types for all 39 objects in the ExoGRAVITY Spectral Library using spectral type templates from the SpeX Prism Libraries \citep{burgasser2014}. We also compute \textit{K}-band spectral indices probing the methane, carbon-monoxide, and water features in the GRAVITY spectra, finding that our ExoGRAVITY sample of bound substellar companions agrees well with field dwarfs from the literature. Most of the objects in the ExoGRAVITY sample are well described in the \textit{K}1-\textit{K}2 color magnitude diagram by a cloudy evolutionary sequence (see Figure~\ref{fig:color_magnitude}), with the exception of HD~4113~c, 51~Eri~b, and HD~13724~b which were all identified as T-type dwarfs with methane-dominated photospheres. Hence, cloudy atmospheres are very common among our current sample of substellar companions observed with GRAVITY. However, we note that this might be due to an observational bias since T-type objects typically have lower effective temperatures and are thus harder to detect with current facilities.

We use the GRAVITY \textit{K}-band spectra to derive bulk properties and isochronal masses for the objects in the ExoGRAVITY Spectral Library. The GRAVITY spectra alone yield reasonable constraints on the objects' bolometric luminosities, consistent within the uncertainties with literature values computed using a broader wavelength coverage. This is because the \textit{K}-band is a powerful spectral range for constraining the atmospheric carbon chemistry of substellar companions which helps to determine whether a given object can be best described by models for clear or cloudy atmospheres. This in turn shrinks the range of luminosities covered by the atmospheric model fitting posteriors and therefore yields tighter luminosity constraints with a typical uncertainty of $\pm0.15$~dex (including systematic errors). Combining the luminosity constraints from GRAVITY with age constraints from the literature where available, we derive isochronal masses for 34 of the 39 objects in the ExoGRAVITY Spectral Library. These evolutionary model masses are in agreement with dynamical masses from the literature for most objects. A notable exception is HD~4113~c which is part of the class of potentially underluminous substellar companions which might indicate the presence of a binary companion as already noted by \citet{cheetham2018}.

\begin{acknowledgements}
    This research has made use of the Jean-Marie Mariotti Center \texttt{Aspro}\footnote{Available at \url{http://www.jmmc.fr/aspro}} service \citep{haubois2014}.
    This work has made use of data from the European Space Agency (ESA) mission {\it Gaia} (\url{https://www.cosmos.esa.int/gaia}), processed by the {\it Gaia} Data Processing and Analysis Consortium (DPAC, \url{https://www.cosmos.esa.int/web/gaia/dpac/consortium}). Funding for the DPAC has been provided by national institutions, in particular the institutions participating in the {\it Gaia} Multilateral Agreement.
    This research has made use of the SVO Filter Profile Service "Carlos Rodrigo", funded by MCIN/AEI/10.13039/501100011033/ through grant PID2023-146210NB-I00.
    This research has benefitted from the SpeX Prism Spectral Libraries, maintained by Adam Burgasser at \url{https://cass.ucsd.edu/ ajb/browndwarfs/spexprism/index.html}.
    This research has benefitted from the Montreal Brown Dwarf and Exoplanet Spectral Library, maintained by Jonathan Gagn\'e.
    S.~L. acknowledges the support of the French Agence Nationale de la Recherche (ANR-21-CE31-0017, ExoVLTI) and of the European Union (ERC, Advanced Grant 101142746, PLANETES).
    J.~J.~W. and A.~C. are supported by NASA XRP Grant 80NSSC23K0280.
\end{acknowledgements}

% WARNING
%-------------------------------------------------------------------
% Please note that we have included the references to the file aa.dem in
% order to compile it, but we ask you to:
%
% - use BibTeX with the regular commands:
%   \bibliographystyle{aa} % style aa.bst
%   \bibliography{Yourfile} % your references Yourfile.bib
%
% - join the .bib files when you upload your source files
%-------------------------------------------------------------------

% \bibliographystyle{aa} 
\bibliographystyle{aa_url}
\vspace{-1cm}
\bibliography{references}

\begin{appendix}

\section{Per epoch GRAVITY spectra}
\label{sec:per_epoch_gravity_spectra}

\begin{figure*}
    \centering
    \includegraphics[trim={0 0 0 0},clip,width=0.98\textwidth]{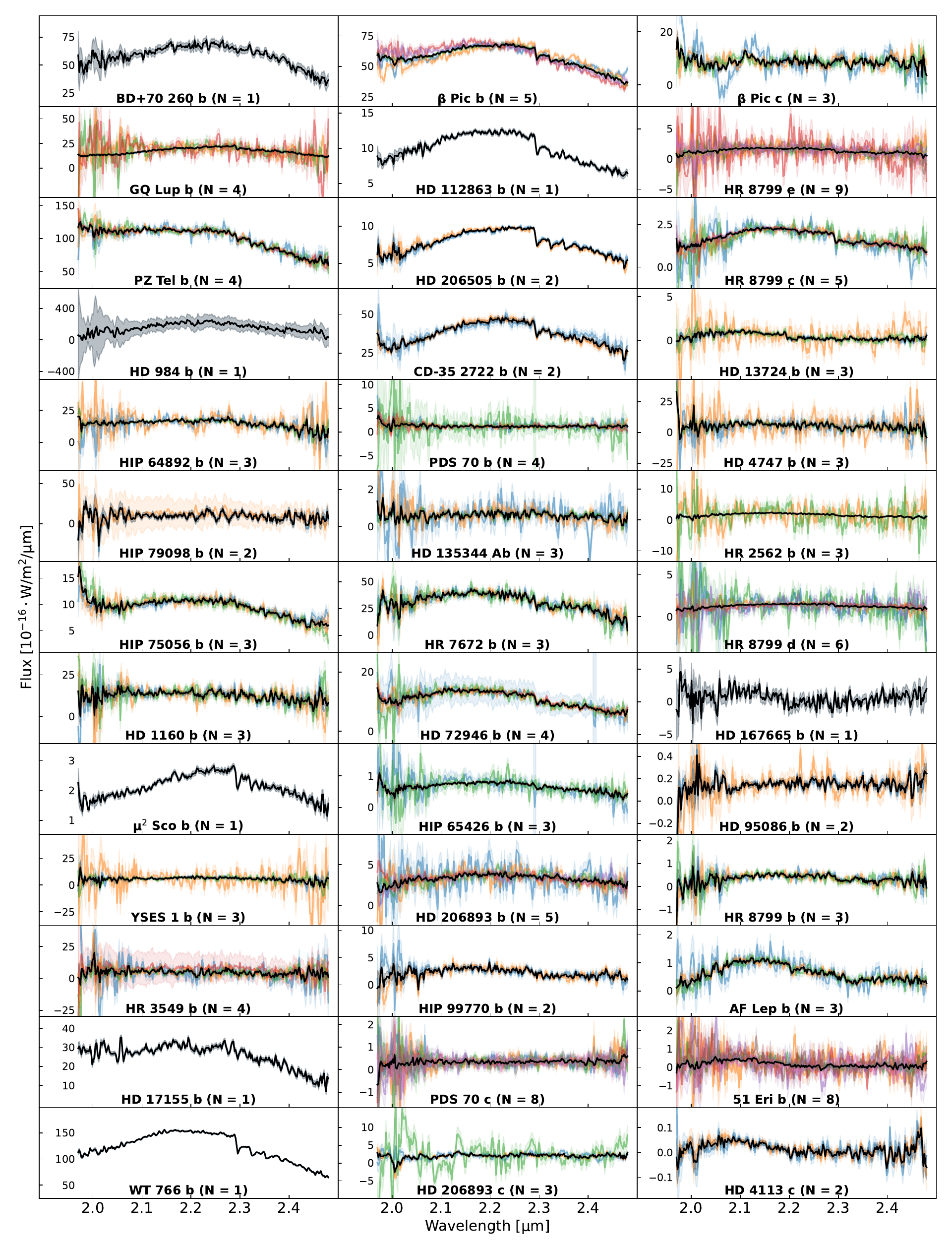}
    \caption{Per epoch GRAVITY spectra (in color, scaled according to the last column of the observations table on Zenodo) and the final combined spectrum (black) for all 39 substellar companions in the ExoGRAVITY Spectral Library. The $N$ epochs are sorted chronologically, starting with the oldest, using the following color palette: blue = 1, orange = 2, green = 3, red = 4, purple = 5, brown = 6, pink = 7, gray = 8, olive = 9.}
    \label{fig:all_spectra}
\end{figure*}

\red{Figure~\ref{fig:all_spectra} presents the per epoch GRAVITY spectra of all 39 substellar companions in the ExoGRAVITY Spectral Library. The shown spectra have already been scaled to bring them into mutual agreement, and the final combined GRAVITY spectrum for each object is shown in black.}

\section{Flux calibration with on-axis host star reference}
\label{sec:flux_calibration_with_on-axis_host_star_reference}

A model spectrum of the flux reference source is needed to convert the GRAVITY contrast spectra into companion flux spectra. We hence derived model spectra for each flux reference source by fitting a grid of stellar model atmospheres to archival \emph{Gaia} \citep{gaia2023}, \emph{Tycho} \citep{hog2000}, \emph{2MASS} \citep{skrutskie2006}, and \emph{WISE} \citep{wright2010} photometry and the \emph{Gaia} XP spectrum \citep{carrasco2021,deangeli2023} where available using the atmospheric modeling toolkit \texttt{species} \citep{stolker2020}. Photometric outliers were identified visually and manually excluded from the fits. For single stars, \emph{WISE}/W3 and \emph{WISE}/W4 photometry were always excluded to avoid any confusion from circumstellar dust or debris often found around young exoplanet host stars \citep[e.g.,][]{smith1984,su2009}. The exception is WT~766, a mature star with no substantial amounts of circumstellar material for which the inclusion of the \emph{WISE}/W3 and \emph{WISE}/W4 photometry lead to a significant improvement of the derived stellar parameters. We also excluded data at wavelengths $<0.37~\micron$ and $>1.01~\micron$ from the \emph{Gaia} XP spectra due to their large uncertainties. Table~\ref{tab:stellar_model_atmospheres} summarizes which photometry was included for which star and whether a \emph{Gaia} XP spectrum was available.

\begin{table*}
\caption{Summary of the archival photometry and spectra included in the atmospheric model fit for each star.}
\label{tab:stellar_model_atmospheres}
\centering
\begin{tabular}{c c c c c c c c c c c c c c c}
\hline\hline
\multirow{2}{*}{Star} & \multicolumn{5}{c}{\emph{Gaia}} & \multicolumn{2}{c}{\emph{Tycho}} & \multicolumn{3}{c}{\emph{2MASS}} & \multicolumn{4}{c}{\emph{WISE}}\\
& XP & $G$ & $G_\text{BP}$ & $G_\text{RP}$ & $G_\text{RVS}$ & $B$ & $V$ & $J$ & $H$ & $K\text{s}$ & $W1$ & $W2$ & $W3$ & $W4$\\
\hline
BD+70 260 & \ding{51} & \ding{51} & \ding{51} & \ding{51} & \ding{55} & \ding{51} & \ding{51} & \ding{51} & \ding{51} & \ding{51} & \ding{51} & \ding{51} & \ding{55} & \ding{55}\\
51 Eri & \ding{51} & \ding{51} & \ding{51} & \ding{51} & \ding{55} & \ding{51} & \ding{51} & \ding{51} & \ding{51} & \ding{51} & \ding{51} & \ding{55} & \ding{55} & \ding{55}\\
AF Lep & \ding{51} & \ding{51} & \ding{51} & \ding{51} & \ding{51} & \ding{51} & \ding{51} & \ding{51} & \ding{51} & \ding{51} & \ding{51} & \ding{55} & \ding{55} & \ding{55}\\
$\beta$ Pictoris & \ding{51} & \ding{51} & \ding{51} & \ding{51} & \ding{51} & \ding{51} & \ding{51} & \ding{51} & \ding{51} & \ding{51} & \ding{51} & \ding{55} & \ding{55} & \ding{55}\\
CD-35 2722 & \ding{51} & \ding{51} & \ding{51} & \ding{51} & \ding{51} & \ding{51} & \ding{51} & \ding{51} & \ding{51} & \ding{51} & \ding{51} & \ding{51} & \ding{55} & \ding{55}\\
GQ Lup & \ding{55} & \ding{51} & \ding{51} & \ding{51} & \ding{55} & \ding{55} & \ding{55} & \ding{51} & \ding{51} & \ding{51} & \ding{55} & \ding{55} & \ding{55} & \ding{55}\\
HD 984 & \ding{55} & \ding{51} & \ding{51} & \ding{51} & \ding{51} & \ding{51} & \ding{51} & \ding{51} & \ding{51} & \ding{51} & \ding{51} & \ding{51} & \ding{55} & \ding{55}\\
HD 1160 & \ding{51} & \ding{51} & \ding{51} & \ding{51} & \ding{51} & \ding{51} & \ding{51} & \ding{51} & \ding{51} & \ding{51} & \ding{51} & \ding{51} & \ding{55} & \ding{55}\\
HD 4113 & \ding{51} & \ding{51} & \ding{51} & \ding{51} & \ding{51} & \ding{51} & \ding{51} & \ding{51} & \ding{51} & \ding{51} & \ding{51} & \ding{51} & \ding{55} & \ding{55}\\
HD 4747 & \ding{51} & \ding{51} & \ding{51} & \ding{51} & \ding{51} & \ding{51} & \ding{51} & \ding{51} & \ding{51} & \ding{51} & \ding{51} & \ding{55} & \ding{55} & \ding{55}\\
HD 13724 & \ding{51} & \ding{51} & \ding{51} & \ding{51} & \ding{51} & \ding{51} & \ding{51} & \ding{51} & \ding{51} & \ding{51} & \ding{51} & \ding{51} & \ding{55} & \ding{55}\\
HD 17155 & \ding{51} & \ding{51} & \ding{51} & \ding{51} & \ding{51} & \ding{51} & \ding{51} & \ding{51} & \ding{51} & \ding{51} & \ding{51} & \ding{51} & \ding{55} & \ding{55}\\
HD 72946 & \ding{51} & \ding{51} & \ding{51} & \ding{51} & \ding{51} & \ding{51} & \ding{55} & \ding{51} & \ding{51} & \ding{51} & \ding{51} & \ding{51} & \ding{55} & \ding{55}\\
HD 95086 & \ding{51} & \ding{51} & \ding{51} & \ding{51} & \ding{51} & \ding{51} & \ding{51} & \ding{51} & \ding{51} & \ding{51} & \ding{51} & \ding{51} & \ding{55} & \ding{55}\\
HD 112863 & \ding{51} & \ding{51} & \ding{51} & \ding{51} & \ding{51} & \ding{51} & \ding{51} & \ding{51} & \ding{51} & \ding{51} & \ding{51} & \ding{51} & \ding{55} & \ding{55}\\
HD 167665 & \ding{55} & \ding{51} & \ding{51} & \ding{51} & \ding{51} & \ding{51} & \ding{51} & \ding{51} & \ding{51} & \ding{51} & \ding{51} & \ding{55} & \ding{55} & \ding{55}\\
HD 206505 & \ding{51} & \ding{51} & \ding{51} & \ding{51} & \ding{51} & \ding{51} & \ding{51} & \ding{51} & \ding{51} & \ding{51} & \ding{51} & \ding{51} & \ding{55} & \ding{55}\\
HD 206893 & \ding{51} & \ding{51} & \ding{51} & \ding{51} & \ding{51} & \ding{51} & \ding{51} & \ding{51} & \ding{51} & \ding{51} & \ding{51} & \ding{51} & \ding{55} & \ding{55}\\
HD 217519 & \ding{51} & \ding{51} & \ding{51} & \ding{51} & \ding{51} & \ding{51} & \ding{51} & \ding{51} & \ding{51} & \ding{51} & \ding{51} & \ding{51} & \ding{55} & \ding{55}\\
HIP 64892 & \ding{51} & \ding{51} & \ding{51} & \ding{51} & \ding{55} & \ding{51} & \ding{51} & \ding{51} & \ding{51} & \ding{51} & \ding{51} & \ding{51} & \ding{55} & \ding{55}\\
HIP 65426 & \ding{51} & \ding{51} & \ding{51} & \ding{51} & \ding{51} & \ding{51} & \ding{51} & \ding{51} & \ding{51} & \ding{51} & \ding{51} & \ding{51} & \ding{55} & \ding{55}\\
HIP 75056 & \ding{51} & \ding{51} & \ding{51} & \ding{51} & \ding{55} & \ding{51} & \ding{51} & \ding{51} & \ding{51} & \ding{51} & \ding{51} & \ding{51} & \ding{55} & \ding{55}\\
HIP 79098 & \ding{51} & \ding{51} & \ding{51} & \ding{51} & \ding{55} & \ding{51} & \ding{51} & \ding{51} & \ding{51} & \ding{51} & \ding{51} & \ding{51} & \ding{55} & \ding{55}\\
HIP 99770 & \ding{51} & \ding{51} & \ding{51} & \ding{51} & \ding{51} & \ding{51} & \ding{51} & \ding{51} & \ding{51} & \ding{51} & \ding{51} & \ding{55} & \ding{55} & \ding{55}\\
HR 2562 & \ding{51} & \ding{51} & \ding{51} & \ding{51} & \ding{51} & \ding{51} & \ding{51} & \ding{51} & \ding{51} & \ding{51} & \ding{51} & \ding{55} & \ding{55} & \ding{55}\\
HR 3549 & \ding{51} & \ding{51} & \ding{51} & \ding{51} & \ding{51} & \ding{51} & \ding{51} & \ding{51} & \ding{51} & \ding{51} & \ding{51} & \ding{51} & \ding{55} & \ding{55}\\
HR 7672 & \ding{51} & \ding{51} & \ding{51} & \ding{51} & \ding{51} & \ding{51} & \ding{51} & \ding{51} & \ding{51} & \ding{51} & \ding{51} & \ding{55} & \ding{55} & \ding{55}\\
HR 8799 & \ding{55} & \ding{51} & \ding{51} & \ding{51} & \ding{51} & \ding{51} & \ding{51} & \ding{51} & \ding{51} & \ding{51} & \ding{51} & \ding{55} & \ding{55} & \ding{55}\\
$\mu^2$ Sco & \ding{51} & \ding{51} & \ding{51} & \ding{51} & \ding{55} & \ding{51} & \ding{51} & \ding{51} & \ding{51} & \ding{51} & \ding{51} & \ding{55} & \ding{55} & \ding{55}\\
PDS 70 & \ding{55} & \ding{51} & \ding{51} & \ding{51} & \ding{51} & \ding{55} & \ding{55} & \ding{51} & \ding{51} & \ding{51} & \ding{51} & \ding{51} & \ding{55} & \ding{55}\\
PZ Tel & \ding{51} & \ding{51} & \ding{51} & \ding{51} & \ding{51} & \ding{51} & \ding{51} & \ding{51} & \ding{51} & \ding{51} & \ding{51} & \ding{51} & \ding{55} & \ding{55}\\
HD 135344 A & \ding{51} & \ding{51} & \ding{51} & \ding{51} & \ding{51} & \ding{51} & \ding{51} & \ding{51} & \ding{51} & \ding{51} & \ding{55} & \ding{55} & \ding{55} & \ding{55}\\
WT 766 & \ding{51} & \ding{55} & \ding{55} & \ding{55} & \ding{55} & \ding{55} & \ding{55} & \ding{51} & \ding{51} & \ding{51} & \ding{51} & \ding{51} & \ding{51} & \ding{51}\\
YSES 1 & \ding{51} & \ding{51} & \ding{51} & \ding{51} & \ding{55} & \ding{51} & \ding{51} & \ding{51} & \ding{51} & \ding{51} & \ding{51} & \ding{51} & \ding{55} & \ding{55}\\
\hdashline
HD 25535 & \ding{51} & \ding{55} & \ding{55} & \ding{55} & \ding{55} & \ding{55} & \ding{55} & \ding{51} & \ding{51} & \ding{51} & \ding{51} & \ding{55} & \ding{51} & \ding{51}\\
HD 30003 & \ding{51} & \ding{55} & \ding{55} & \ding{55} & \ding{55} & \ding{55} & \ding{55} & \ding{55} & \ding{55} & \ding{55} & \ding{51} & \ding{55} & \ding{51} & \ding{51}\\
HD 174536 & \ding{55} & \ding{51} & \ding{51} & \ding{51} & \ding{51} & \ding{51} & \ding{51} & \ding{51} & \ding{51} & \ding{51} & \ding{51} & \ding{51} & \ding{55} & \ding{55}\\
HD 123227 & \ding{55} & \ding{55} & \ding{55} & \ding{55} & \ding{55} & \ding{55} & \ding{55} & \ding{51} & \ding{51} & \ding{51} & \ding{51} & \ding{51} & \ding{51} & \ding{51}\\
HD 91881 & \ding{55} & \ding{55} & \ding{55} & \ding{55} & \ding{55} & \ding{55} & \ding{55} & \ding{51} & \ding{51} & \ding{51} & \ding{51} & \ding{55} & \ding{51} & \ding{51}\\
HD 73900 & \ding{51} & \ding{55} & \ding{55} & \ding{55} & \ding{55} & \ding{51} & \ding{51} & \ding{51} & \ding{51} & \ding{51} & \ding{51} & \ding{51} & \ding{51} & \ding{51}\\
HD 196885 & \ding{51} & \ding{51} & \ding{51} & \ding{51} & \ding{51} & \ding{51} & \ding{51} & \ding{51} & \ding{51} & \ding{51} & \ding{51} & \ding{51} & \ding{55} & \ding{55}\\
HR 5362 & \ding{55} & \ding{55} & \ding{55} & \ding{55} & \ding{55} & \ding{55} & \ding{55} & \ding{51} & \ding{51} & \ding{51} & \ding{51} & \ding{51} & \ding{51} & \ding{51}\\
\hline
\end{tabular}
\tablefoot{\red{Objects below the horizontal dashed line are binary stars (swap calibrators).}}
\end{table*}

The archival spectrophotometry of all single stars were then fitted with the BT-NextGen grid of stellar model atmospheres from \citet{allard2012}. This grid has three free parameters, effective temperature $T_\text{eff} \in [2600, 30000]$~K, surface gravity $\log g \in [3, 6]$, and metallicity $[\text{Fe}/\text{H}] \in [0, 0.5]$~dex. Besides, we also fitted for the radius of the star $R \in [0.1, 40]~\text{R}_\odot$ and the parallax $\pi$ for which we used a Gaussian prior based on the \emph{Gaia} EDR3 parallax and its uncertainties \citep{gaia2020}. To avoid degeneracies between the inferred stellar parameters with the limited spectral coverage and resolution in the archival spectrophotometry, we fixed the surface gravity for all stars at the value reported in the Gaia DR3 catalog \citep{gaia2023}\footnote{With the exception of a few stars for which no surface gravity or extinction value was available in the Gaia DR3 catalog, see Table~\ref{tab:stellar_parameters_single}} and the metallicity at the solar value of $[\text{Fe}/\text{H}] = 0$~dex. We note that literature values for the metallicity often vary significantly based on different assumptions even for many of the well-studied exoplanet host stars considered here so that setting the stellar metallicity based on these literature values is difficult. Hence, and given that the metallicity distribution of young stars in the solar neighborhood has been shown to cluster around the solar value \citep[e.g.,][]{haywood2001}, we consider $[\text{Fe}/\text{H}] = 0$~dex a conservative choice. For HD~217519 and HIP~99770, no surface gravity constraints were available in the literature. For these two stars, we used additional Gaussian priors on the radius of HD~217519 of $15.1\pm0.5~\text{R}_\odot$ based on \citet{cruzalebes2019} and on the mass of HIP~99770 of $1.8\pm0.2~\text{M}_\odot$ based on \citet{currie2023} and left the surface gravity as a free parameter. Moreover, we applied the ISM extinction law from \citet{cardelli1989} with the extinction parameter $A_V$ set to the value of the $A_0$ parameter in the \emph{Gaia} DR3 catalog \citep{gaia2023}${}^\text{5}$.

\begin{figure}
    \centering
    \includegraphics[width=\linewidth]{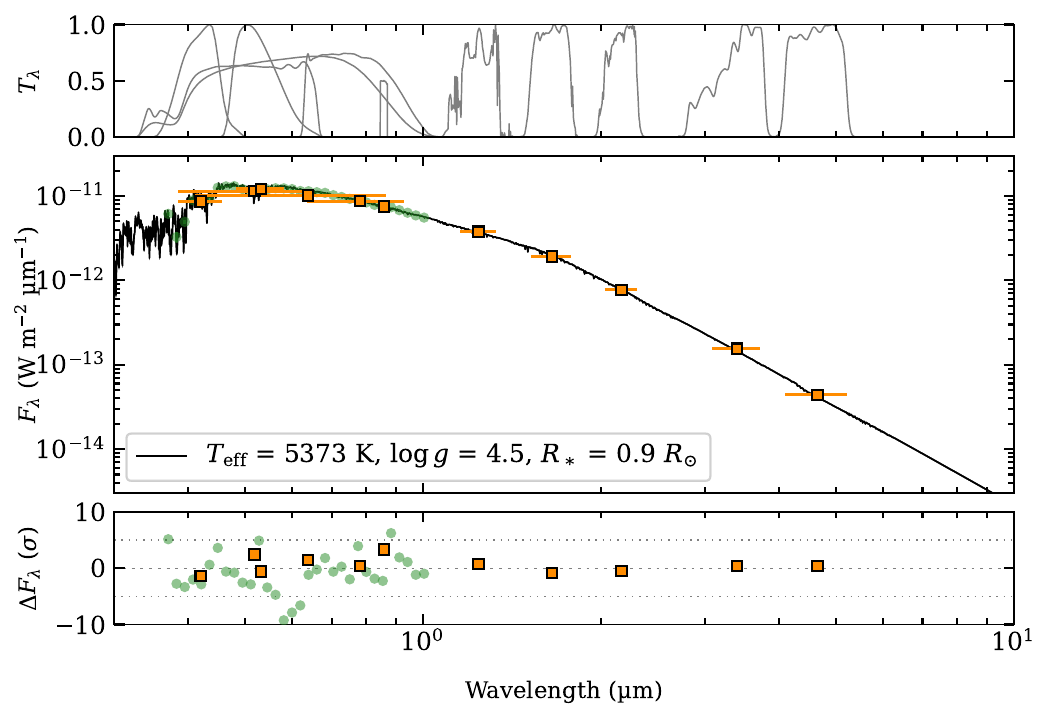}
    \caption{Stellar model atmosphere for HD~206505~A (black line). The thin gray lines are 30 samples drawn randomly from the posterior distribution (almost invisible here because the posterior is tightly constrained). The photometry and the \emph{Gaia} XP spectrum included in the fit are shown in orange and green, respectively. For better readability, only every 10th data point of the \emph{Gaia} XP spectrum is shown. The top panel shows the filter transmission curve for each photometric point and the bottom panel shows the residuals between the data and the best fit stellar model atmosphere.}
    \label{fig:hd206505_spectrum}
\end{figure}

For GQ~Lup and PDS~70, we additionally fitted for a second blackbody component to describe the circumstellar dust emission which becomes evident in their spectral energy distributions in the \emph{2MASS} and \emph{WISE} passbands. The effective temperature and radius of this second blackbody component were bounded to $T_\text{eff,disk} \in [10, 2000]$~K and $R_\text{disk} \in [10, 1000]~\text{R}_\text{J}$. The best fit stellar model parameters were then inferred with \texttt{species} using nested sampling with \texttt{PyMultiNest} \citep{buchner2014}. Table~\ref{tab:stellar_parameters_single} presents the stellar model parameters inferred from the atmospheric model fitting process. For all stars except HIP~99770, the archival \emph{2MASS} $K\text{s}$-band photometry lies within $3\sigma$ of the best fit stellar model spectrum. The best fit effective temperature is typically within $\sim200$~K of literature values where available, except for HIP~79098. For this particularly hot star ($T_\text{eff} > 10000$~K), our inferred effective temperature differs from literature values by $\sim600$~K. An example stellar model atmosphere for HD~206505 is shown in Figure~\ref{fig:hd206505_spectrum}.

\begin{table*}
\caption{Stellar parameters for the on-axis host stars inferred from the atmospheric model fitting.}
\label{tab:stellar_parameters_single}
\centering
\begin{tabular}{c c c c c c c}
\hline\hline
Star & $T_\text{eff}$ [K] & $\log g$ & $R$ [$\text{R}_\odot$] & $\pi$ [mas] & $d$ [pc] & $A_V$ [mag]\\
\hline
BD+70 260 & $6242^{+21}_{-20}$ & 4.07${}^\text{(a)}$ & $1.47^{+0.01}_{-0.01}$ & $8.52^{+0.01}_{-0.01}$ & $117.3^{+0.1}_{-0.1}$ & 0.0${}^\text{(a)}$\\
51 Eri & $7386^{+20}_{-17}$ & 4.14${}^\text{(a)}$ & $1.48^{+0.01}_{-0.01}$ & $33.44^{+0.06}_{-0.05}$ & $29.9^{+0.0}_{-0.1}$ & 0.0${}^\text{(a)}$\\
AF Lep & $6062^{+39}_{-37}$ & 4.15${}^\text{(a)}$ & $1.21^{+0.01}_{-0.01}$ & $37.25^{+0.01}_{-0.01}$ & $26.8^{+0.0}_{-0.0}$ & 0.0${}^\text{(a)}$\\
$\beta$ Pic & $8053^{+37}_{-36}$ & 4.21${}^\text{(a)}$ & $1.53^{+0.01}_{-0.01}$ & $50.93^{+0.10}_{-0.10}$ & $19.6^{+0.0}_{-0.0}$ & 0.0${}^\text{(a)}$\\
CD-35 2722 & $3713^{+7}_{-7}$ & 4.64${}^\text{(a)}$ & $0.58^{+0.00}_{-0.00}$ & $44.72^{+0.01}_{-0.01}$ & $22.4^{+0.0}_{-0.0}$ & 0.0${}^\text{(a)}$\\
GQ Lup & $4306^{+35}_{-35}$ & 3.70${}^\text{(b)}$ & $1.75^{+0.06}_{-0.07}$ & $6.49^{+0.02}_{-0.02}$ & $154.1^{+0.5}_{-0.5}$ & 0.4${}^\text{(a)}$\\
HD 984 & $6268^{+29}_{-27}$ & 4.21${}^\text{(a)}$ & $1.21^{+0.01}_{-0.01}$ & $21.88^{+0.02}_{-0.02}$ & $45.7^{+0.0}_{-0.0}$ & 0.0${}^\text{(a)}$\\
HD 1160 & $10386^{+46}_{-66}$ & 4.50${}^\text{(c)}$ & $1.67^{+0.01}_{-0.01}$ & $8.27^{+0.02}_{-0.03}$ & $120.9^{+0.4}_{-0.3}$ & 0.2${}^\text{(a)}$\\
HD 4113 & $5700^{+22}_{-21}$ & 4.36${}^\text{(a)}$ & $1.07^{+0.01}_{-0.01}$ & $23.83^{+0.02}_{-0.02}$ & $42.0^{+0.0}_{-0.0}$ & 0.0${}^\text{(a)}$\\
HD 4747 & $5438^{+23}_{-23}$ & 4.50${}^\text{(a)}$ & $0.76^{+0.01}_{-0.01}$ & $53.05^{+0.02}_{-0.02}$ & $18.8^{+0.0}_{-0.0}$ & 0.0${}^\text{(a)}$\\
HD 13724 & $5842^{+13}_{-12}$ & 4.31${}^\text{(a)}$ & $1.05^{+0.00}_{-0.01}$ & $23.02^{+0.01}_{-0.01}$ & $43.4^{+0.0}_{-0.0}$ & 0.0${}^\text{(a)}$\\
HD 17155 & $4792^{+11}_{-10}$ & 4.61${}^\text{(d)}$ & $0.71^{+0.00}_{-0.00}$ & $35.61^{+0.07}_{-0.06}$ & $28.1^{+0.0}_{-0.1}$ & 0.1${}^\text{(h)}$\\
HD 72946 & $5656^{+23}_{-22}$ & 4.42${}^\text{(a)}$ & $0.92^{+0.01}_{-0.01}$ & $38.98^{+0.03}_{-0.03}$ & $25.7^{+0.0}_{-0.0}$ & 0.0${}^\text{(a)}$\\
HD 95086 & $7623^{+33}_{-26}$ & 4.18${}^\text{(a)}$ & $1.48^{+0.01}_{-0.01}$ & $11.57^{+0.01}_{-0.01}$ & $86.5^{+0.1}_{-0.1}$ & 0.0${}^\text{(a)}$\\
HD 112863 & $5329^{+15}_{-14}$ & 4.52${}^\text{(a)}$ & $0.78^{+0.00}_{-0.01}$ & $26.95^{+0.02}_{-0.02}$ & $37.1^{+0.0}_{-0.0}$ & 0.0${}^\text{(a)}$\\
HD 167665 & $6228^{+30}_{-27}$ & 4.23${}^\text{(a)}$ & $1.30^{+0.01}_{-0.01}$ & $32.40^{+0.06}_{-0.07}$ & $30.9^{+0.1}_{-0.1}$ & 0.0${}^\text{(a)}$\\
HD 206505 & $5373^{+13}_{-13}$ & 4.45${}^\text{(a)}$ & $0.89^{+0.01}_{-0.00}$ & $22.77^{+0.01}_{-0.01}$ & $43.9^{+0.0}_{-0.0}$ & 0.0${}^\text{(a)}$\\
HD 206893 & $6554^{+29}_{-27}$ & 4.13${}^\text{(a)}$ & $1.31^{+0.01}_{-0.01}$ & $24.53^{+0.03}_{-0.02}$ & $40.8^{+0.0}_{-0.0}$ & 0.0${}^\text{(a)}$\\
HD 217519 & $4741^{+12}_{-11}$ & $3.07^{+0.11}_{-0.05}$ & $17.25^{+0.16}_{-0.17}$ & $1.73^{+0.01}_{-0.01}$ & $576.7^{+3.7}_{-3.8}$ & 0.4${}^\text{(h)}$\\
HIP 64892 & $10600^{+120}_{-99}$ & 4.24${}^\text{(a)}$ & $1.74^{+0.01}_{-0.02}$ & $8.36^{+0.03}_{-0.03}$ & $119.6^{+0.5}_{-0.5}$ & 0.0${}^\text{(i)}$\\
HIP 65426 & $8746^{+19}_{-18}$ & 4.23${}^\text{(a)}$ & $1.71^{+0.01}_{-0.01}$ & $9.30^{+0.02}_{-0.02}$ & $107.5^{+0.3}_{-0.3}$ & 0.0${}^\text{(i)}$\\
HIP 75056 & $8231^{+43}_{-29}$ & 4.22${}^\text{(e)}$ & $1.61^{+0.01}_{-0.02}$ & $8.20^{+0.03}_{-0.03}$ & $121.9^{+0.4}_{-0.4}$ & 0.2${}^\text{(e)}$\\
HIP 79098 & $11074^{+100}_{-85}$ & 3.76${}^\text{(a)}$ & $3.77^{+0.08}_{-0.08}$ & $6.48^{+0.13}_{-0.13}$ & $154.4^{+3.1}_{-3.0}$ & 0.3${}^\text{(a)}$\\
HIP 99770 & $8060^{+34}_{-33}$ & $4.08^{+0.05}_{-0.07}$ & $1.97^{+0.01}_{-0.01}$ & $24.55^{+0.06}_{-0.06}$ & $40.7^{+0.1}_{-0.1}$ & 0.0${}^\text{(j)}$\\
HR 2562 & $6615^{+20}_{-26}$ & 4.22${}^\text{(a)}$ & $1.40^{+0.01}_{-0.01}$ & $29.47^{+0.01}_{-0.01}$ & $33.9^{+0.0}_{-0.0}$ & 0.0${}^\text{(a)}$\\
HR 3549 & $10449^{+100}_{-49}$ & 4.24${}^\text{(a)}$ & $2.06^{+0.01}_{-0.01}$ & $10.55^{+0.03}_{-0.03}$ & $94.8^{+0.2}_{-0.2}$ & 0.1${}^\text{(a)}$\\
HR 7672 & $6002^{+25}_{-20}$ & 4.37${}^\text{(a)}$ & $1.05^{+0.01}_{-0.01}$ & $56.27^{+0.03}_{-0.03}$ & $17.8^{+0.0}_{-0.0}$ & 0.0${}^\text{(a)}$\\
HR 8799 & $7339^{+29}_{-27}$ & 4.07${}^\text{(a)}$ & $1.45^{+0.01}_{-0.01}$ & $24.46^{+0.03}_{-0.03}$ & $40.9^{+0.1}_{-0.1}$ & 0.0${}^\text{(a)}$\\
$\mu^2$ Sco & $21608^{+467}_{-321}$ & 3.71${}^\text{(a)}$ & $5.90^{+0.09}_{-0.09}$ & $6.18^{+0.09}_{-0.10}$ & $161.9^{+2.8}_{-2.3}$ & 0.2${}^\text{(a)}$\\
PDS 70 & $4247^{+41}_{-39}$ & 4.15${}^\text{(a)}$ & $1.14^{+0.03}_{-0.03}$ & $8.90^{+0.01}_{-0.01}$ & $112.4^{+0.2}_{-0.2}$ & 0.3${}^\text{(a)}$\\
PZ Tel & $5263^{+16}_{-16}$ & 4.25${}^\text{(a)}$ & $1.22^{+0.01}_{-0.01}$ & $21.16^{+0.02}_{-0.02}$ & $47.3^{+0.0}_{-0.0}$ & 0.0${}^\text{(a)}$\\
HD 135344 A & $10003^{+93}_{-91}$ & 3.80${}^\text{(f)}$ & $1.46^{+0.02}_{-0.01}$ & $7.41^{+0.03}_{-0.03}$ & $134.9^{+0.5}_{-0.5}$ & 0.3${}^\text{(f)}$\\
WT 766 & $3237^{+6}_{-6}$ & 4.55${}^\text{(g)}$ & $0.25^{+0.00}_{-0.00}$ & $74.45^{+0.54}_{-0.52}$ & $13.4^{+0.1}_{-0.1}$ & 0.1${}^\text{(h)}$\\
YSES 1 & $4952^{+15}_{-15}$ & 4.41${}^\text{(a)}$ & $0.99^{+0.01}_{-0.01}$ & $10.61^{+0.01}_{-0.01}$ & $94.2^{+0.1}_{-0.1}$ & 0.3${}^\text{(a)}$\\
\hline
\end{tabular}
\tablefoot{(a) \citealt{gaia2023}, (b) \citealt{donati2012}, (c) \citealt{mesa2020}, (d) \citealt{barbato2023}, (e) \citealt{bochanski2018}, (f) \citealt{stolker2025}, (g) \citealt{steinmetz2020}, (h) \citealt{gaia2018}, (i) \citealt{chen2012}, (j) \citealt{murpy2017}.}
\end{table*}

\begin{table*}
\caption{Stellar parameters for the swap binary star calibrators inferred from the atmospheric model fitting.}
\label{tab:stellar_parameters_binary}
\centering
\begin{tabular}{c c c c c c c c c c c}
\hline\hline
Star & $T_\text{eff,A}$ [K] & $\log g_\text{A}$ & $R_\text{A}$ [$\text{R}_\odot$] & $T_\text{eff,B}$ [K] & $\log g_\text{B}$ & $R_\text{B}$ [$\text{R}_\odot$] & $\pi$ [mas] & $d$ [pc] & $A_V$ [mag] & $c_{K\text{s}}$\\
\hline
HD 25535 & $5872^{+69}_{-70}$ & $4.15^{+0.03}_{-0.03}$ & $1.50^{+0.03}_{-0.03}$ & $5650^{+555}_{-482}$ & $4.36^{+0.06}_{-0.05}$ & $1.14^{+0.07}_{-0.07}$ & $21.66^{+0.01}_{-0.01}$ & $46.2^{+0.0}_{-0.0}$ & 0.0${}^\text{(a)}$ & 0.56\\
HD 30003 & $5580^{+1}_{-1}$ & $4.32^{+0.04}_{-0.04}$ & $1.14^{+0.03}_{-0.03}$ & $5589^{+20}_{-24}$ & $4.42^{+0.04}_{-0.04}$ & $1.01^{+0.03}_{-0.03}$ & $32.42^{+0.01}_{-0.01}$ & $30.8^{+0.0}_{-0.0}$ & 0.0${}^\text{(b)}$ & 0.80\\
HD 174536 & $4811^{+30}_{-30}$ & $3.00^{+0.00}_{-0.00}$ & $35.2^{+1.0}_{-1.0}$ & $1496^{+83}_{-108}$ & $3.00^{+0.00}_{-0.00}$ & $35.1^{+2.3}_{-2.4}$ & $1.51^{+0.03}_{-0.04}$ & $663^{+17}_{-15}$ & 1.3${}^\text{(a)}$ & 0.03\\
HD 123227 & $6115^{+623}_{-646}$ & $4.18^{+0.06}_{-0.06}$ & $1.51^{+0.10}_{-0.09}$ & $5968^{+790}_{-753}$ & $4.25^{+0.08}_{-0.08}$ & $1.36^{+0.11}_{-0.10}$ & $22.23^{+0.05}_{-0.05}$ & $45.0^{+0.1}_{-0.1}$ & 0.0${}^\text{(b)}$ & 0.79\\
HD 91881 & $6117^{+6}_{-6}$ & $4.16^{+0.03}_{-0.03}$ & $1.56^{+0.02}_{-0.02}$ & $6385^{+1015}_{-723}$ & $4.50^{+0.08}_{-0.07}$ & $0.97^{+0.07}_{-0.08}$ & $23.17^{+0.02}_{-0.02}$ & $43.2^{+0.0}_{-0.0}$ & 0.0${}^\text{(a)}$ & 0.41\\
HD 73900 & $6884^{+36}_{-40}$ & $4.27^{+0.04}_{-0.05}$ & $1.45^{+0.02}_{-0.02}$ & $5636^{+47}_{-51}$ & $4.54^{+0.06}_{-0.07}$ & $0.87^{+0.03}_{-0.03}$ & $24.41^{+0.03}_{-0.03}$ & $41.0^{+0.1}_{-0.1}$ & 0.0${}^\text{(b)}$ & 0.29\\
HD 196885 & $6267^{+18}_{-16}$ & $4.28^{+0.03}_{-0.03}$ & $1.38^{+0.01}_{-0.01}$ & $3549^{+121}_{-124}$ & $4.58^{+0.06}_{-0.06}$ & $0.57^{+0.04}_{-0.04}$ & $29.41^{+0.02}_{-0.02}$ & $34.0^{+0.0}_{-0.0}$ & 0.0${}^\text{(a)}$ & 0.06\\
HR 5362 & $5784^{+30}_{-36}$ & $3.00^{+0.00}_{-0.00}$ & $11.77^{+0.06}_{-0.05}$ & $5947^{+71}_{-73}$ & $3.84^{+0.68}_{-0.52}$ & $2.02^{+0.08}_{-0.08}$ & $8.07^{+0.02}_{-0.03}$ & $124^{+0}_{-0}$ & 0.0${}^\text{(b)}$ & 0.03\\
\hline
\end{tabular}
\tablefoot{\red{The SPHERE IRDIS $K\text{s}$-band contrast estimated from the GRAVITY data is given by $c_{K\text{s}}$.} (a) \citealt{gaia2023}, (b) set to 0 since no extinction estimates available in the literature -- we note that this value is consistent with other targets at similar distances in our stellar database.}
\end{table*}

\begin{figure}
    \centering
    \includegraphics[width=\linewidth]{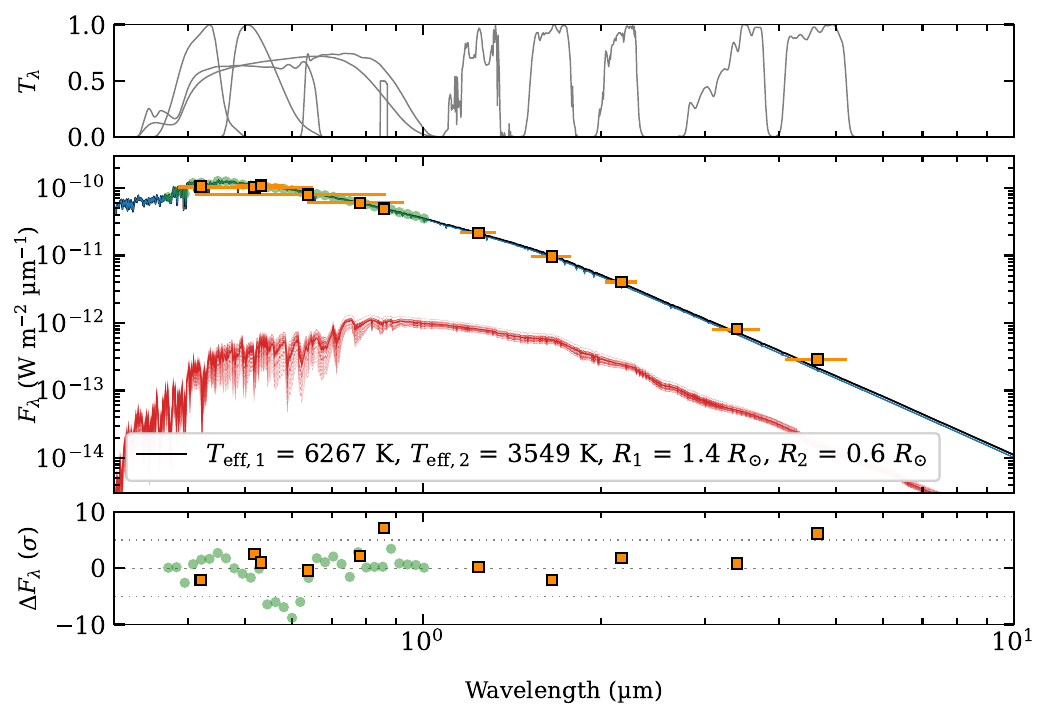}
    \caption{Stellar model atmospheres for HD~196885~A and~B (blue and red line). The combined spectrum is shown in black. The thin light lines are 30 samples drawn randomly from the posterior distribution. The photometry and the \emph{Gaia} XP spectrum included in the fit are shown in orange and green, respectively. For better readability, only every 10th data point of the \emph{Gaia} XP spectrum is shown. The top panel shows the filter transmission curve for each photometric point and the bottom panel shows the residuals between the data and the best fit combined model atmosphere.}
    \label{fig:hd196885_spectrum}
\end{figure}

To obtain the final companion flux spectra $F_\text{comp}$, the companion contrast spectra $c_\text{comp}$ measured by GRAVITY were then multiplied with the host star model spectra $F_\text{star}$ evaluated at the same wavelength nodes as the GRAVITY spectra, that is
\begin{align}
    F_\text{comp} = c_\text{comp} \cdot F_\text{star}.
\end{align}
The covariances of the companion contrast spectra output by the GRAVITY pipeline and the uncertainties of the stellar model spectra obtained as the standard deviation over 100 model spectra drawn randomly from the posterior distribution were propagated accordingly. We note that the uncertainties reported for the derived stellar parameters only reflect the statistical uncertainties from the fit, but ignore any systematic uncertainties in the underlying archival spectrophotometry and in the interpolation of the stellar model grid. This has a negligible impact on our final uncertainties, however, because the uncertainties propagated from the GRAVITY pipeline are at least an order of magnitude larger than those propagated from the stellar model spectrum.

\section{Flux calibration with swap binary star reference}
\label{sec:flux_calibration_with_off-axis_binary_star_reference}

The swap binary star calibrators observed before or after the companion off-axis observations have the primary purpose of calibrating the OPD between the two off-axis fields created by the roof prism. This is required to measure accurate companion astrometry. In some cases, a dedicated on-axis observation with the science spectrometer on the companion host star was taken before or after the companion off-axis observation which can then be used to construct a companion contrast spectrum relative to the host star. In all other cases, this is unfortunately not possible and the companion contrast spectrum was calibrated with respect to the swap binary star calibrator. This means that a model spectrum of the binary star (i.e., both of its components) needed to be derived for such cases. This process is complicated since most of the used binary stars (unlike the nearby exoplanet host stars) are rather poorly studied systems with little (resolved) archival spectrophotometry available.

For the binary stars, we used the same grid of stellar model atmospheres as for the single stars, except for HD~174536 and HD~196885. For these two binary stars, the effective temperature of the secondary component converged towards the lower grid boundary of 2600~K, so that another grid of stellar model atmospheres had to be consulted. We decided to use the BT-Settl (CIFIST) grid \citep{allard2012} which has only two free parameters, effective temperature $T_\text{eff} \in [1200, 7000]$~K and surface gravity $\log g \in [3, 5.5]$. Consistent with the fits for single stars, this grid has been computed at solar metallicity $[\text{Fe}/\text{H}] = 0$~dex. In addition to the previously described parameters fitted for the single stars, we also computed synthetic SPHERE IRDIS $K\text{s}$-band contrasts between the primary and secondary components of the binary stars from the GRAVITY observations and included them in the \texttt{species} fits. Given the large observed variations in the slope of the observed GRAVITY contrast spectra for individual binary stars, we decided to first compute an average contrast spectrum from all available epochs and then combine it into a single photometric contrast data point. For this, the SPHERE IRDIS $K\text{s}$-band filter was an ideal choice because it covers a large fraction of the GRAVITY bandpass. A systematic error of 10\% was associated with these SPHERE IRDIS $K\text{s}$-band contrasts. Additional priors used for each individual binary star system can be found in the next paragraph. Table~\ref{tab:stellar_parameters_binary} contains the extinction values from the \emph{Gaia} catalog used in the fits and the best fit stellar model parameters inferred from the atmospheric model fitting process. An example stellar model atmosphere for HD~196885~AB is shown in Figure~\ref{fig:hd196885_spectrum}.

\noindent\textbf{HD~25535:} This binary star has an angular separation of $\sim1\arcsec$ \citep{nowak2024} and is resolved by \emph{Gaia}. There is an effective temperature estimate for the primary star available in the \emph{Gaia} catalog \citep{gaia2023} so that we used a Gaussian prior of $T_\text{eff,A} = 5876\pm99$~K in our fit. In addition, \citet{makarov2021} found a mass ratio of $0.941\pm0.002$ between the primary and secondary component of the binary from \emph{Hipparcos} and \emph{Gaia} astrometry and \citet{kervella2019} reported a mass for the primary component of $M_\text{A} = 1.153\pm0.058~\text{M}_\odot$ which we also included as Gaussian priors in our fit.

\noindent\textbf{HD~30003:} This binary star has an angular separation of $\sim4\arcsec$ and is resolved by \emph{Gaia}. There is an effective temperature estimate for both the primary and the secondary star available in the \emph{Gaia} catalog \citep{gaia2023}. However, we only used a Gaussian prior of $T_\text{eff,A} = 5580\pm2$~K in our fit. We did not include a prior on the effective temperature of the secondary component of the binary because the combined Gaia XP spectrum strongly preferred a $\sim180$~K lower effective temperature for the secondary component than what is reported in the \emph{Gaia} catalog. In addition, \citet{makarov2021} found a mass ratio of $0.989\pm0.003$ between the primary and secondary component of the binary from \emph{Hipparcos} and \emph{Gaia} astrometry and \citet{tokovinin2014} reported masses for the primary and secondary component of $M_\text{A} = 1.03\pm0.10~\text{M}_\odot$ and $M_\text{B} = 1.01\pm0.10~\text{M}_\odot$ which we also included as Gaussian priors in our fit.

\noindent\textbf{HD~174536:} This binary star has an angular separation of $\sim1.5\arcsec$ \citep{nowak2024} and is resolved by \emph{Gaia}. There is an effective temperature estimate for the primary star available in the \emph{Gaia} catalog \citep{gaia2023} so that we used a Gaussian prior of $T_\text{eff,A} = 4848\pm293$~K in our fit. We note that HD~174536 has been classified as a giant star by \citet{houk1988} and our binary star model fit converged to radii of $R_\text{A} = 35.2~\text{R}_\odot$ and $R_\text{B} = 35.1~\text{R}_\odot$ which appears to confirm this finding. However, the stellar models that we used are not specifically made to describe giant stars which might have consequences for our binary star model of HD~174536 and the GQ~Lup~b off-axis epoch for which HD~174536 was used as an amplitude reference.

\noindent\textbf{HD~123227:} This binary star has an angular separation of $\sim1\arcsec$ \citep{nowak2024} and is resolved by \emph{Gaia}. There is no effective temperature estimate for either of its two components available in the \emph{Gaia} catalog so that we let both $T_\text{eff,A}$ and $T_\text{eff,B}$ vary freely in our fit. However, \citet{tokovinin2014} reported masses for the primary and secondary component of $M_\text{A} = 1.27\pm0.10~\text{M}_\odot$ and $M_\text{B} = 1.21\pm0.10~\text{M}_\odot$ which we included as Gaussian priors in our fit.

\noindent\textbf{HD~91881:} This binary star has an angular separation of about $\sim1.3\arcsec$ \citep{nowak2024} and is resolved by \emph{Gaia}. There is an effective temperature estimate for the primary star available in the \emph{Gaia} catalog \citep{gaia2023} so that we used a Gaussian prior of $T_\text{eff,A} = 6117\pm9$~K in our fit. In addition, \citet{makarov2021} found a mass ratio of $0.857\pm0.002$ between the primary and secondary component of the binary from \emph{Hipparcos} and \emph{Gaia} astrometry and \citet{tokovinin2014} reported masses for the primary and secondary component of $M_\text{A} = 1.31\pm0.10~\text{M}_\odot$ and $M_\text{B} = 1.07\pm0.10~\text{M}_\odot$ which we also included as Gaussian priors in our fit.

\noindent\textbf{HD~73900:} This binary star has an angular separation of $\sim1\arcsec$ \citep{nowak2024} and is resolved by \emph{Gaia}. There is no effective temperature estimate for either of its two components available in the \emph{Gaia} catalog. However, \citet{abu-dhaim2022} reported effective temperatures and masses for the primary and secondary component of $T_\text{eff,A} = 6880\pm70$~K and $T_\text{eff,B} = 5630\pm70$~K and $M_\text{A} = 1.44\pm0.15~\text{M}_\odot$ and $M_\text{B} = 0.98\pm0.13~\text{M}_\odot$ which we included as Gaussian priors in our fit.

\noindent\textbf{HD~196885:} This binary star has an angular separation of $\sim0.5\arcsec$ \citep{nowak2024} and is \emph{not} resolved by \emph{Gaia}. However, based on the findings of \citet{chauvin2023}, we used Gaussian priors of $T_\text{eff,A} = 6340\pm39$~K and $T_\text{eff,B} = 3660\pm190$~K for the effective temperature and $M_\text{A} = 1.3\pm0.1~\text{M}_\odot$ and $M_\text{B} = 0.45\pm0.01~\text{M}_\odot$ for the mass of the primary and secondary component, respectively. Here, we converted the spectral type constraint of M1$\pm$1 for the secondary component from \citet{chauvin2023} to an effective temperature constraint of $3660\pm190$~K using \citet{pecaut2013}.

\noindent\textbf{HR~5362:} This binary star has an angular separation of $\sim3.5\arcsec$ and is resolved by \emph{Gaia}. There is no effective temperature estimate for either of its two components available in the \emph{Gaia} catalog. However, \citet{corbally1984} reported spectral types of G8 III and G0 IV for the primary and secondary component, respectively. Here, we converted these spectral type constraints to effective temperature constraints of $T_\text{eff,A} = 5480\pm100$~K and $T_\text{eff,B} = 5930\pm120$~K using \citet{pecaut2013} and used them as Gaussian priors in our fit. Our binary star model fit converged to radii of $R_\text{A} = 11.8~\text{R}_\odot$ and $R_\text{B} = 2.0~\text{R}_\odot$ which appears to confirm their classification as giant and subgiant star from \citet{corbally1984}. However, the stellar models that we used are not specifically made to describe giant stars which might have consequences for our binary star model of HR~5362 and the three YSES~1~b off-axis epochs for which HR~5362 was used as an amplitude reference.

The final companion flux spectra were then obtained as
\begin{align}
    F_\text{companion} = c_\text{companion} \cdot \langle F_\text{star}\rangle \cdot \sqrt{\frac{F_\text{A} \cdot F_\text{B}}{\langle F_\text{A}\rangle \cdot \langle F_\text{B}\rangle}},
\end{align}
where $\langle F\rangle$ denotes the flux average over the GRAVITY bandpass and the subscripts A and B denote the binary star A and B components. We note that $\langle F_\text{star}\rangle$ can always be obtained from the low-resolution spectrum of the host star recorded simultaneously by the GRAVITY fringe tracker, even during off-axis observations. The uncertainties of the stellar model spectra were obtained similar as in the single star case and propagated accordingly together with the covariances of the companion contrast spectra. In a few cases, two binary stars were available and used for the flux calibration to improve the SNR. The final companion flux spectra were then obtained as
\begin{align}
    F_\text{companion} = c_\text{companion} \cdot \langle F_\text{star}\rangle \cdot \sqrt[4]{\frac{F_\text{A} \cdot F_\text{B} \cdot F_\text{C} \cdot F_\text{D}}{\langle F_\text{A}\rangle \cdot \langle F_\text{B}\rangle \cdot \langle F_\text{C}\rangle \cdot \langle F_\text{D}\rangle}},
\end{align}
where the subscripts C and D denote the binary star A and B components of the second binary star.

\section{Determination of spectral types and uncertainties}
\label{sec:determination_of_spectral_types_and_uncertainties}

\begin{figure*}
    \centering
    \includegraphics[trim={0 0 0 0},clip,width=0.97\textwidth]{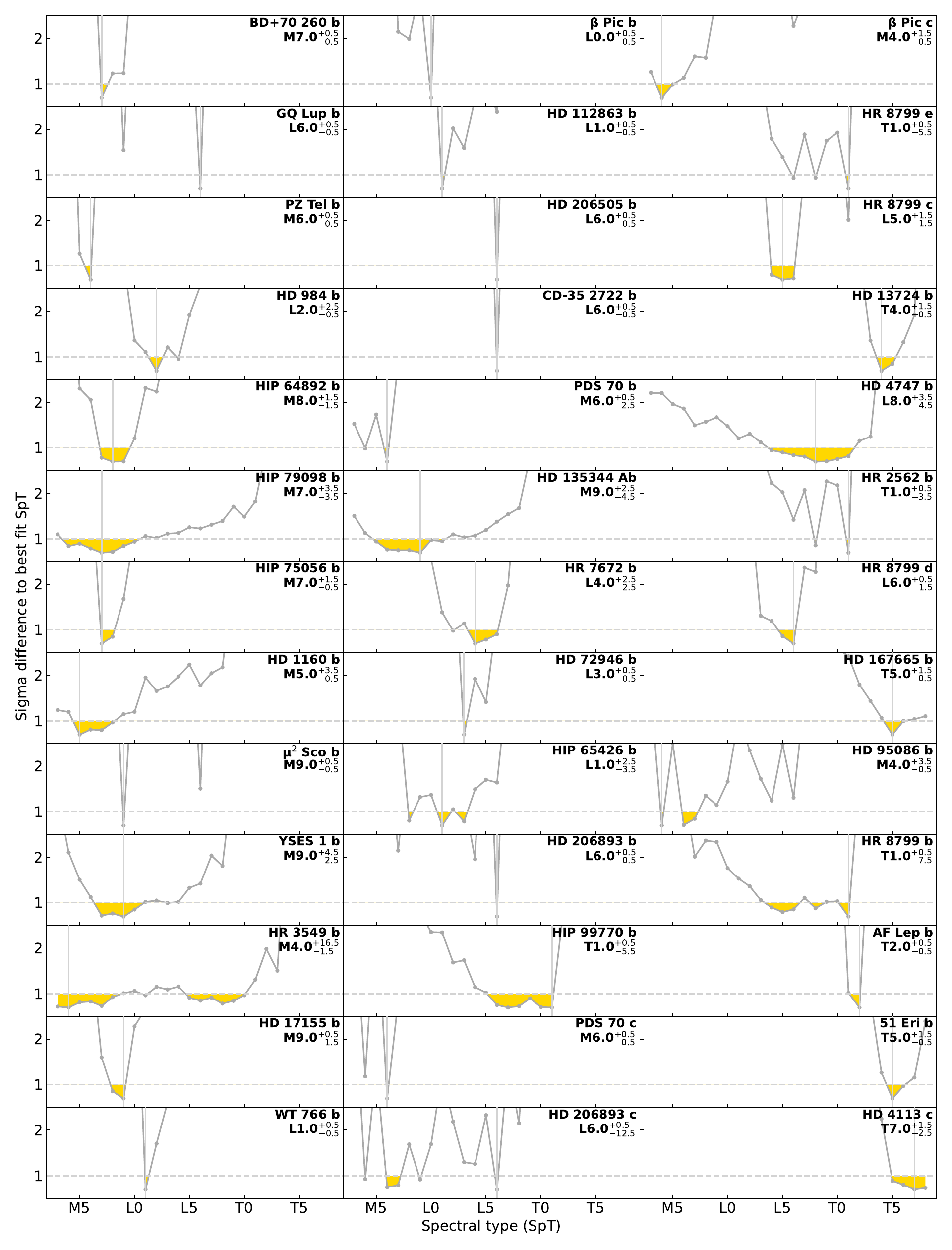}
    \caption{Determination of the spectral type and its uncertainties for all 39 substellar companions in the ExoGRAVITY Spectral Library. The dark gray lines show the sigma difference between the given spectral type and the best fitting spectral type, the latter of which is also highlighted by a vertical light gray line. The spectral type range within $1\sigma$ of the best fitting spectral type (area highlighted in gold) is considered the spectral type uncertainty.}
    \label{fig:all_spectral_types}
\end{figure*}

\red{Figure~\ref{fig:all_spectral_types} shows how the spectral type and its uncertainties were derived for each of the 39 objects in the ExoGRAVITY Spectral Library. The range of spectral types within $1\sigma$ of the best fit is shown in gold.}

\section{Posteriors from the atmospheric model fitting}
\label{sec:posteriors_from_the_atmospheric_model_fitting}

\begin{figure*}
    \centering
    \includegraphics[trim={0 0.5cm 0 0.3cm},clip,width=0.94\textwidth]{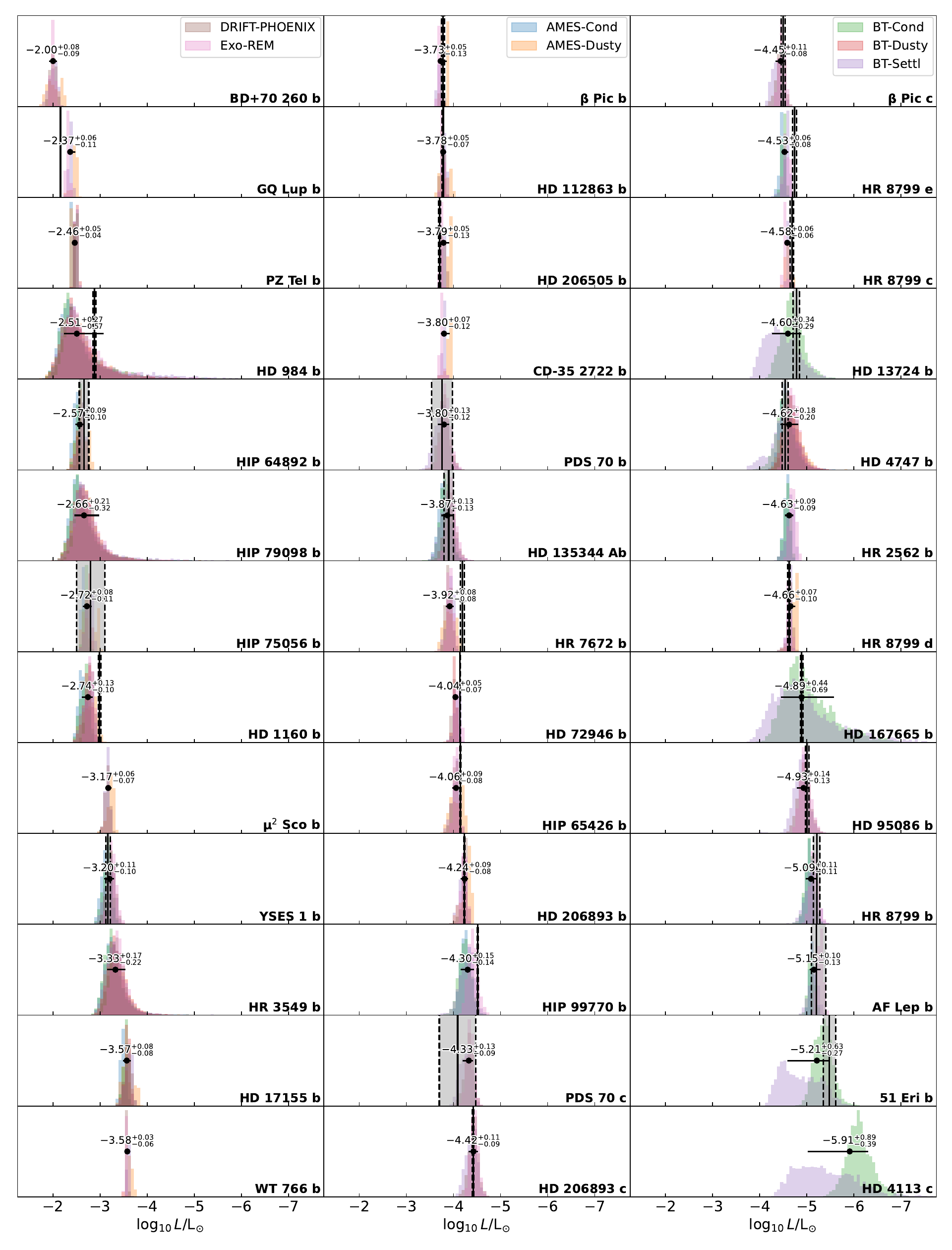}
    \caption{Posterior distribution of the bolometric luminosity obtained from fitting the GRAVITY spectra of the 39 objects in the ExoGRAVITY Spectral Library with different atmospheric model grids. For each object, only the posteriors from the grids that provide a reduced $\chi^2$ value close to the one of the best-fitting model are shown. The median value and its uncertainties (16th and 84th percentiles) over all shown posteriors are shown by a black data point. We note that an additional systematic uncertainty of 10\% ($\sim0.05$~dex) should be added in quadrature to the GRAVITY-derived luminosity when using it for further analyses to account for flux calibration errors. Where available, the literature values from Table~\ref{tab:companion_parameters} are shown by a black vertical line with a gray shaded area \red{(bordered by dashed black lines)} for the uncertainties.}
    \label{fig:all_lum_posteriors}
\end{figure*}

\red{Figure~\ref{fig:all_lum_posteriors} shows the posterior distributions of the bolometric luminosity from the atmospheric model fits of all 39 objects in the ExoGRAVITY Spectral Library. The posteriors obtained with different atmospheric model grids are shown in different colors. Where available, luminosity constraints from the literature are shown for reference.}

\end{appendix}

\end{document}